\documentclass[thmsa,12pt]{article}
\normalfont\sffamily
\usepackage{amsfonts}
\usepackage{amssymb}
\usepackage[active]{srcltx}
\usepackage{color}
\usepackage[x11names]{xcolor}
\usepackage{amsmath}
\usepackage{amsthm}
\usepackage{bm} 
\usepackage{graphicx}
\usepackage{textcomp}
\usepackage{a4wide}
\usepackage{chicago}
\usepackage{multirow}
\usepackage{diagbox}
\usepackage{color}
\usepackage[format=hang]{caption}
\usepackage{subcaption}
\usepackage{url}
\usepackage{xfrac}
\usepackage{lscape}
\usepackage{longtable}

\usepackage[normalem]{ulem}
\usepackage{comment}
\usepackage{pxfonts}
\usepackage[margin=3cm]{geometry} 
\usepackage{titlesec}

\usepackage{setspace}
\usepackage{footmisc}
\usepackage{accents}
\usepackage{appendix}

\usepackage{array,booktabs,ragged2e}
\newcolumntype{R}[1]{>{\RaggedRight}p{#1}}

\usepackage{cellspace}
\usepackage{hyperref}

\def \be {\begin{equation}}
\def \ee {\end{equation}}

\begin{document}

\def\title #1{\begin{center}
{\Large {\sc #1}}
\end{center}}
\def\author #1{\begin{center} {#1}
\end{center}}

\def\KMN{{KMN }}

\setstretch{1.1}

\begin{titlepage}
    \phantomsection \label{Titlepage}
    \addcontentsline{toc}{section}{Title page}

\renewcommand{\thefootnote}{\fnsymbol{footnote}}\addtocounter{footnote}{1}
\title{\sc 
The Art and Beauty of Voting Power\\ \medskip \  }
\author{Sascha Kurz\\ {\small Dept.\ of Mathematics, University of Bayreuth, Germany\\sascha.kurz@uni-bayreuth.de  }}

\author{Alexander Mayer\\ {\small Dept.\ of Economics, University of Bayreuth, Germany\\alexander.mayer@uni-bayreuth.de  }}

\author{Stefan Napel\\ {\small Dept.\ of Economics, University of Bayreuth, Germany\\stefan.napel@uni-bayreuth.de }}



\begin{center} {\bf {\sc Abstract}} \end{center}
{\small 
We exhibit the hidden beauty of weighted voting and voting power by applying a generalization of the Penrose-Banzhaf index to social choice rules.
Three players who have multiple votes in a committee decide between three options by plurality rule, Borda's rule, antiplurality rule, or one of the many scoring rules in between.
A~priori influence on outcomes is quantified in terms of how players' probabilities to be pivotal for the committee decision compare to a dictator.
The resulting numbers are represented in triangles that map out structurally equivalent voting weights.
Their geometry and color variation reflect fundamental differences between voting rules, such as their inclusiveness and transparency.
} 

\vspace{0.2cm}

\begin{description}
{\small
\item[Keywords:]
weighted voting $\cdot$ weighted committee games $\cdot$ scoring rules $\cdot$ simple voting games $\cdot$ collective choice 
\item[JEL codes:] D71 $\cdot$ C71 $\cdot$ C63 
}
\end{description}

\vspace{1.6cm}

\vfill

\noindent {\footnotesize We thank an anonymous referee for his or her thoughtful reading and several helpful suggestions.}

\end{titlepage}

\addtocounter{footnote}{-1}

\setstretch{1.25}

\pagenumbering{arabic}


\section{Introduction}\label{sec:Introduction}

Individual voting rights entail a potential to affect collective decisions. 
Greater numbers of votes controlled by large shareholders, party leaders, delegates to a council or committee, etc.\ typically increase the respective influence. 
It is not trivial, though, to tell how much greater the influence of, e.g., a voter wielding 25\% of all votes is compared to one wielding only 5\%. 
This holds even when an a~priori perspective is adopted, meaning that one purposely leaves aside personal affiliations between the voters and empirical preference information. 
Various indices try to rigorously quantify voting power 
in order to address this problem.
 
For binary collective decision making of the \emph{yes}-or-\emph{no} kind~-- formalized by simple voting games (cf.\ \citeNP[Ch.~10]{vonNeumann/Morgenstern:1953} or \citeNP{Taylor/Zwicker:1999})~-- prominent examples of voting power indices are the Penrose-Banzhaf, Shapley-Shubik, and Holler-Packel indices (cf.\ \citeNP{Penrose:1946}, \citeNP{Banzhaf:1965}, \citeNP{Shapley/Shubik:1954}, \citeNP{Holler/Packel:1983}). 
Some of them have been extended to non-binary settings such as the determination of a winner from a set of more than two options by alternative methods of social choice (cf.\  \citeNP{Kurz/Mayer/Napel:2021:Influence}, for instance).
The respective winner could be a particular law selected from multiple legal drafts, the managing director of the IMF chosen from a shortlist of three nominees, a presidential candidate who is picked from various primary contenders, and so on. 

Applied to a particular voting body such as a parliament or party convention, the IMF Executive Board, the EU~Council or the US~Electoral College,\footnote{We recommend the contributions in \citeN{Holler/Nurmi:2013} for a good overview of typical applications of power indices. Somewhat atypical applications are discussed by \citeN{Kovacic/Zoli:2021} and \citeN{Napel/Welter:2021}. \citeN{Napel:2018} provides a short introduction to power measurement with many further references.} power indices illuminate some of the discrete mathematical structure that underlies collective choices.
They identify possible swings between losing and winning coalitions, which make outcomes depend on a voter's behavior, or more generally they measure potential variation of the winning candidate that derives from a single voter's input to the decision process. 
Light is cast on either a specific institutional arrangement, when the focus is on influence of distinct members of a voting body relative to another, or on multiple competing arrangements.
One may evaluate, for instance, the implications of a change of the majority threshold or a switch from plurality voting to a runoff system (e.g., see \citeNP{Maskin/Sen:2016} on plurality voting in US presidential primaries).

So power indices yield insights with political or economic meaning. 
Common questions are:
To what extent can a given shareholder control a corporation and may, perhaps, be held responsibility for its actions?
Is the voting power of two parties at least approximately proportional to their seat shares in parliament?
Is a given allocation of voting rights to delegates from different constituencies (e.g., US states in the Electoral College, member countries in the EU~Council, departments in a university senate, etc.) `fair' under a particular set of normative premises? Etcetera. 

This article, however, is \emph{not} pursuing serious questions of any such kind. 
We here employ a power index for non-binary decisions with a seemingly superficial and primarily visual purpose: we try to convey the hidden beauty of weighted voting and want to exhibit artistic aspects of the power that voters can derive from their voting weights. 

The article's main part therefore consists of several pages of color images. 
Depending on personal taste, they may be of interest and produce enjoyment without any further explanation.
At the same time, they represent the result of hours of computer calculations (several weeks, in fact).
They give a graphical picture of the formal structure behind collective decision making by three players~-- individuals or homogeneous groups of voters~-- on three candidates.

It will be assumed that winners are determined by a plurality vote, an antiplurality vote, or one of the many scoring rules that lie `in between', such as Borda's voting rule. 
Other voting methods like the various rules that focus on pairwise majority comparisons (Copeland's rule, Kemeny-Young rule, etc.) are amenable to the same kind of representation. 
They generate less scintillating results however (cf.\ \citeNP{Kurz/Mayer/Napel:2019:WCG}). 

Appreciators of art and beauty without interest in the formal framework are welcome to jump to Section~\ref{sec:Bilder}.
For all others, we will first provide a short introduction to weighted committee games (Section~\ref{sec:committees_and_scoring_rules}).\footnote{
See  \shortciteN{Kurz/Mayer/Napel:2019:WCG} for details and related literature: the article defines weighted committee games, characterizes and counts equivalence classes for selected voting rules, and provides lists of structurally distinct committees.
\citeN{Mayer/Napel:2021:Scoring} does similarly for the special case of scoring rules.
\shortciteN{Kurz/Mayer/Napel:2021:Influence} generalizes the Penrose-Banzhaf and Shapely-Shubik indices to committee games.
For a practical application of the framework see \citeN{Mayer/Napel:2020}.
} 
These games generalize traditional weighted voting from binary majority decisions to social choice from any finite number of options. 
We explain the pertinent generalization of the Penrose-Banzhaf power index and how this can literally provide colorful insights into how voting weights determine voters' influence on collective decisions (Section~\ref{sec:power_and_color}).
Possible economic and political implications are briefly pointed out in Section~\ref{sec:conclusion}.


\section{Weighted committees and scoring rules}\label{sec:committees_and_scoring_rules}

Binary \emph{weighted voting games} involve a set of $n$ players who respectively wield voting weights denoted by $w_1, w_2, \ldots, w_n\ge 0$. 
The majority threshold or decision quota is set to $q>0$:
the players can jointly pass any proposal that is made to them if the subset of players who support the motion wield a combined voting weight of at least $q$.
For instance, if players~1, 2 and 3 have weights of $w_1=40\%$, $w_2=35\%$ and $w_3=25\%$ and face a quota $q=51\%$ then at least two players must support a proposal for it to pass.
Subsets $S\subseteq N$ of the set $N$ of players with $\sum_{i \in S}w_i\ge q$ are also referred to as winning coalitions, while subsets $T\subseteq N$ with $\sum_{i \in T}w_i< q$ are known as losing coalitions. 

Weighted voting games constitute a special kind of (binary) \emph{simple voting games}. 
The latter do not necessarily require a link between winning or losing to weights and a quota. 
They merely assume that the full player set $N$ is winning, the empty set $\varnothing$ is losing, and winning is monotonic with respect to set inclusion, i.e., any superset of a winning coalition is also winning.

Simple voting games are commonly specified in set-theoretic terms. 
This is done either by directly listing a subclass of winning coalitions (typically those that are minimal with respect to set inclusion) or using an indicator function $v$ that takes a set $S\subseteq N$ as its argument and outputs a 1 if and only if $S$ is winning. 
A simple voting game can, however, also be described as a mapping from the set of all possible profiles of players' preferences over a status quo option $a_1$ and an alternative motion $a_2$ that is voted on to the set of possible outcomes, specifying for each preference profile the collective decision $a_1$ or $a_2$ that is adopted. 
\emph{Weighted committee games} follow this route and allow to handle also non-binary decisions.

In particular, the latter consider a finite set $N$ of $n\ge 2$ players and assume that each player~$i\in N$ has strict preferences $P_i $ over a set $A=\{a_1,\dots,a_m\}$ of $m\ge 2$ options that the committee needs to choose from. 
The set of all $m\cdot (m-1)\cdot \ldots \cdot 1=m!$ conceivable strict preference orderings on $A$ is denoted by $\mathcal{P}(A)$.
Any collective decision rule can then be conceived of as a mathematical mapping $\rho\colon\mathcal{P}(A)^n \to A$. 
This translates any preference profile $\mathbf{P}=(P_1,\dots,P_n)$ into a single winning alternative $a^*=\rho(\mathbf{P})$.  
The respective combination $(N, A, \rho)$ of a set of voters, a set of alternatives, and a decision rule is referred to as a committee game or as a \emph{committee} for short. 

A committee $(N, A, \rho)$ is called a \emph{weighted plurality committee} if the decision rule $\rho$ amounts to each voter~$i$ casting $w_i\ge 0$ votes for its favorite option and then selecting the alternative $a^*$ that received the most votes as winner.
Similarly,  for a \emph{weighted antiplurality committee} the decision rule $\rho$ amounts to each voter~$i$ casting $w_i\ge 0$ negative or dissenting votes for its least preferred option and then the alternative $a^*$ that received the fewest dissenting votes becomes the winner. 
In case of ties we suppose that they are resolved lexicographically: if, for instance, $A=\{a_1, a_2, a_3, a_4\}$ and these alternatives respectively receive 3, 4, 0, and 4 plurality votes from $n=11$ voters with a weight of $w_i=1$ each, then $a_2$ rather than $a_4$ is chosen. 
Declaring both $a_2$ and $a_4$ to be winners and tossing a coin to reach a resolute decision would be a possibility too. 
But randomness would complicate the mathematical exposition without changing the illustrations below.

(Weighted) plurality and antiplurality committees are special cases of (weighted) \emph{scoring committees}. 
These entail the application of a \emph{scoring rule}: the winning candidate or option $a^*$ always is the one that received the highest total score from the voters.
Candidates' scores are determined by their positions in each voter's preference ranking and a given vector $\mathbf{s}=(s_1, s_2, \ldots, s_m)$ with $s_1\ge s_2\ge \ldots \ge s_m$ and $s_1\neq s_m$: when voters' weights are $w_1=\ldots=w_n=1$, any alternative $a\in A$ receives $s_1$ points for every voter that ranks $a$ first, $s_2$ points for every voter that ranks $a$ second, and so on. 
When the voters have non-uniform weights $w_1, \ldots, w_n\ge 0$, the respective points derived from how voter~$i\in N$ ranks the alternatives are multiplied by $w_i$.

For illustration, suppose that a committee~-- perhaps the board of a sports club~-- involves four voter groups, i.e., players $N=\{1,2,3,4\}$, with group~1 wielding 5 votes, group~2 having 4 votes, group~3 wielding 3 votes, and group~4 having only one vote. 
The weights are summarized by $\mathbf{w}=(5,4,3,1)$.
The voters must select one of three candidates, say, Ann, Bob, or Clara, to lead their club. 
\begin{samepage}
	
Let the players' preferences $\mathbf{P}=(P_1,P_2,P_3,P_4)$ rank the candidates as in the following table:
\newpage
\renewcommand{\arraystretch}{1.2}
\begin{table}[h!]
	\begin{center}
		\begin{tabular}{|c|c|c|c|c|}
			\hline
			                   & $P_1$ & $P_2$ & $P_3$ & $P_4$ \\ \hline
			$1^\text{st}$ best &  Bob  &  Ann  &  Ann  &  Bob  \\
			$2^\text{nd}$ best & Clara & Clara &  Bob  & Clara \\
			$3^\text{rd}$ best &  Ann  &  Bob  & Clara &  Ann  \\ \hline
		\end{tabular} 
	\end{center}
\end{table}

\end{samepage}

\noindent Using the scoring vector $\mathbf{s}=(1, 0, 0)$ amounts to a weighted plurality vote: Ann receives a total score of $5\cdot 0+4\cdot 1 + 3\cdot 1 +1\cdot 0=7$; Bob's score is $5\cdot 1+4\cdot 0 + 3\cdot 0 +1\cdot 1=6$; and Clara, being ranked first by nobody, gets a score of 0. 
Ann wins.

Had above committee used the scoring vector $\mathbf{s}=(1, 1, 0)$ instead, Clara would have won with a score of 10 vs.\ 7 for Ann and 9 for Bob.
The latter vector $\mathbf{s}$ is equivalent to conducting an antiplurality vote because minimizing the number of dissenting votes is the same as maximizing the number of non-dissenting votes captured by $\mathbf{s}=(1,1,0)$.

An example of a voting rule in between plurality and antiplurality is \emph{Borda's rule}: 
voters state their full preferences and each candidate $a$ receives as many points from a given voter~$i$ as there are candidates that $i$ ranks below $a$. 
For instance, Bob would receive 2~points for each vote wielded by group~1, 0~points from group~2, 1~point for each of the votes held by group~3, and again 2~points from group~4. This gives Bob a total Borda score of $5\cdot 2+ 4\cdot 0+3\cdot 1+1\cdot 2=15$.
That number is greater than the analogous figures of 14 for Ann and 10 for Clara. 
So Bob would win if scoring vector $\mathbf{s}=(2, 1, 0)$ or Borda's rule were used.

Maximizing the total score given the scoring vector $\mathbf{s}=(2, 1, 0)$ is equivalent to maximizing the total score for vectors $\mathbf{s}'=(1, \sfrac{1}{2}, 0)$ or $\mathbf{s}''=(4, 3, 2)$. 
Vector $\mathbf{s}'$ merely halves above numbers, while preserving the order of Ann's, Bob's and Clara's totals. 
Similarly, using $\mathbf{s}''$ raises all candidates' scores by $(5+4+3+1)\cdot 2=26$ without changing their order.
In particular, scoring winners are invariant to positive affine transformations of the adopted scoring vector $\mathbf{s}$.
Hence, whenever a committee picks a winner from three candidates by a scoring rule~-- plurality, antiplurality, Borda, or any other rule that determines the winner by evaluating the candidates' positions in the applicable preference profile $\mathbf{P}$ with decreasing scores~-- it is without loss of generality to suppose a vector  $\mathbf{s}=(1, s, 0)$ such that $0\le s\le 1$.

When a committee with player set $N$ and voting weights $\mathbf{w}=(w_1, \ldots, w_n)$ decides on a set $A$ of $m=3$ alternatives and uses a decision rule $\rho$ that amounts to applying the scoring vector $\mathbf{s}=(1, s, 0)$, we write $(N,A,r^s|\mathbf{w})$ instead of $(N,A,\rho)$. We refer to such committee as a (weighted) \emph{$s$-scoring committee} (see \citeNP{Mayer/Napel:2021:Scoring}). 

We have seen that the special $s$-scoring committees with $s=1$, $s=\sfrac{1}{2}$, and $s=0$ amount to weighted plurality, Borda, and antiplurality committees. 
As above example illustrates, the respective committees differ for the considered voting weights $\mathbf{w}=(5,4,3,1)$.
Namely, they select a different winner from three candidates for at least some configuration of preferences. 
Similarly, two plurality committees ($s=0$) are different depending on whether weights $\mathbf{w}=(5,4,3,1)$ or weights $\mathbf{w'}=(5,1,1,1)$ apply to the players (club members, shareholders, parties, etc.):
for the profile $\mathbf{P}$ at hand, Bob rather than Ann would be selected if $\mathbf{w}$ were replaced by $\mathbf{w'}$. 

We call two committees $(N, A, \rho)$ and  $(N, A, \rho')$  that never select different winners from set $A$~-- no matter which preference profile $\mathbf{P}=(P_1,\ldots,P_n)$ is considered~-- \emph{equivalent}. This means that the respective mappings $\rho\colon\mathcal{P}(A)^n \to A$ and $\rho'\colon\mathcal{P}(A)^n \to A$ are identical, denoted by $\rho\equiv \rho'$.
We can have $\rho\equiv \rho'$ even though the verbal descriptions of $\rho$ and $\rho'$  may differ. For instance, $\rho$ may be described as plurality voting with weights $\mathbf{w'}=(5,1,1,1)$ and $\rho'$ as dictatorship of voter~1: the committee in either case always chooses the alternative that is ranked first according to $P_1$.

When two $s$-scoring committees  $(N,A,r^s|\mathbf{w})$ and  $(N,A,r^s|\mathbf{w'})$ with $\mathbf{w}\neq  \mathbf{w'}$ are equivalent, i.e., $r^s|\mathbf{w}\equiv r^s|\mathbf{w'}$, we learn that it does not matter which of the two voting weight distributions prevails: decisions will coincide. 
From the perspective of an outsider who does not care about the labeling of the players, this is also true if weights $\mathbf{w''}$ are used that only label the players differently than $\mathbf{w}$. 

Consider, for instance, $\mathbf{w''}=(1,3,4,5)$ instead of $\mathbf{w}=(5,4,3,1)$ in our example. This represents the same abstract decision structure except that player numbers have changed.
In particular, the situation for the preferences $\mathbf{P}=(P_1,P_2,P_3, P_4)$ depicted in the table above for weights $\mathbf{w}$ (with Ann winning under plurality rule, Clara under antiplurality rule, etc.) is the same as that with weights $\mathbf{w''}$ and preferences $\mathbf{P''}=(P_4,P_3,P_2, P_1)$. 
We then say that $\mathbf{w}$ and $\mathbf{w''}$ are \emph{structurally equivalent} under the considered $s$-scoring rule: the implied mappings $r^s|\mathbf{w}$ and $r^s|\mathbf{w''}$ become equivalent after suitably relabeling the players.

Having fixed a scoring rule, such as $r^s$ for $s=1$, the set of all weights $\mathbf{w}=(w_1, \ldots, w_n)$ that are structurally equivalent to a given reference distribution of weights  $\mathbf{\tilde w}=(\tilde w_1, \ldots, \tilde w_n)$ can be grouped together and form an \emph{equivalence class} of weights: 
if two weight distributions $\mathbf{w}\neq \mathbf{w'}$ belong to the same class, the corresponding $s$-scoring committees always produce identical decisions (once labels of the players are harmonized). If the weight distributions belong to different classes, there exists at least some preference configuration $\mathbf{P}$ that results in different committee decisions. 

For antiplurality rule ($s=1$) and three players ($n=3$), it turns out that there are only five different equivalence classes~-- namely those that correspond to reference weights of $\mathbf{\tilde w}=(1,0,0)$, $(1,1,0)$, $(1,1,1)$, $(2,1,1)$ and $(2,2,1)$.
Any other distribution of weights among three players is structurally equivalent to one of these, i.e., leads to the same decisions after suitable relabeling (cf.\ \shortciteNP{Kurz/Mayer/Napel:2019:WCG}).
Similarly, there are only six structurally different plurality committees for three players. 
The respective reference weights equal the five just listed for antiplurality rule in addition to $\mathbf{\tilde w}=(3,2,2)$.

\begin{figure}
	\centering
	\includegraphics[width=0.55\textwidth]{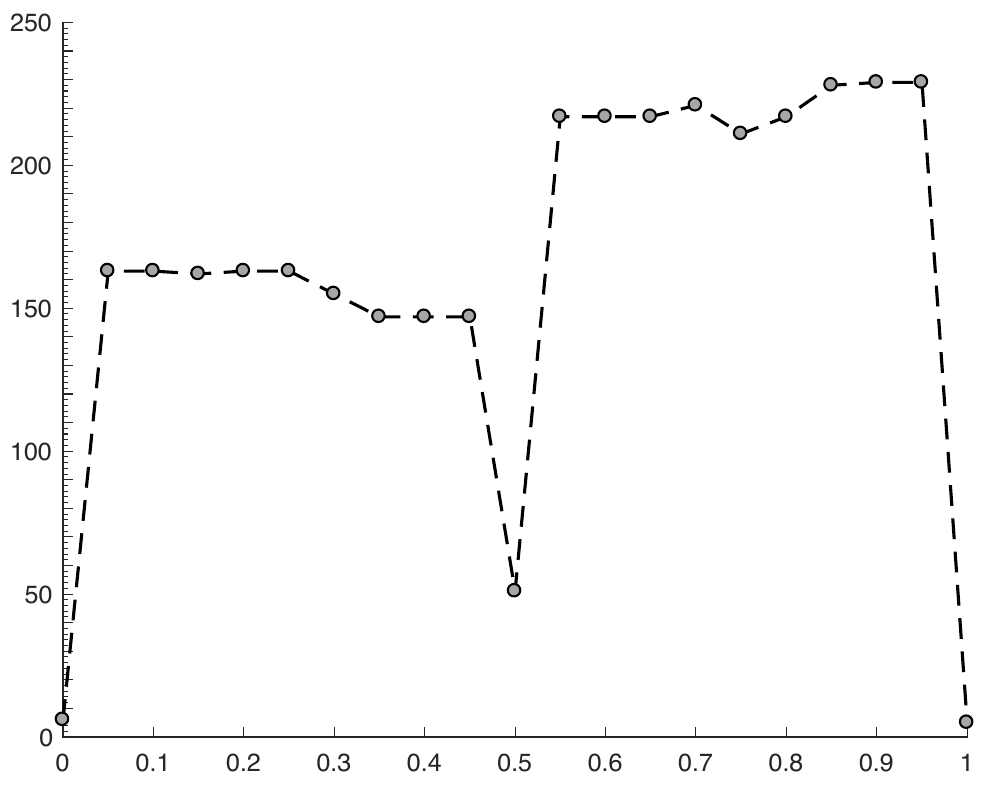}
	\caption{\small Number of $s$-scoring committees for $m=3$ and $s\in\{0,0.05,0.1,\dots,1\}$}
	\label{fig:M-shape}
\end{figure}

The numbers of structurally distinct $s$-scoring committees for $s=\sfrac{1}{2}$ (Borda) and, more pronouncedly, for $0<s<\sfrac{1}{2}$ or $\sfrac{1}{2}<s<1$ are much higher than those for $s=0$ and $s=1$.
Exact values have not been published for all $s$ yet, but \citeN{Mayer/Napel:2021:Scoring} provide the numbers of equivalence classes for all $s$ that are integer multiples of $\sfrac{1}{20}$.
These numbers range up to 229 and exhibit an M-shaped pattern reproduced in Figure~\ref{fig:M-shape}.

Knowing that a given weight distribution among three players structurally amounts to, say, $(2,1,1)$ can simplify the analysis of the respective committee: the distribution of voting power is as if weights were $(2,1,1)$. So are players' manipulation incentives, strategic voting equilibria, the scope for voting paradoxes, etc.

\begin{figure}
 	\begin{center}
 		\includegraphics[width=0.65\textwidth]{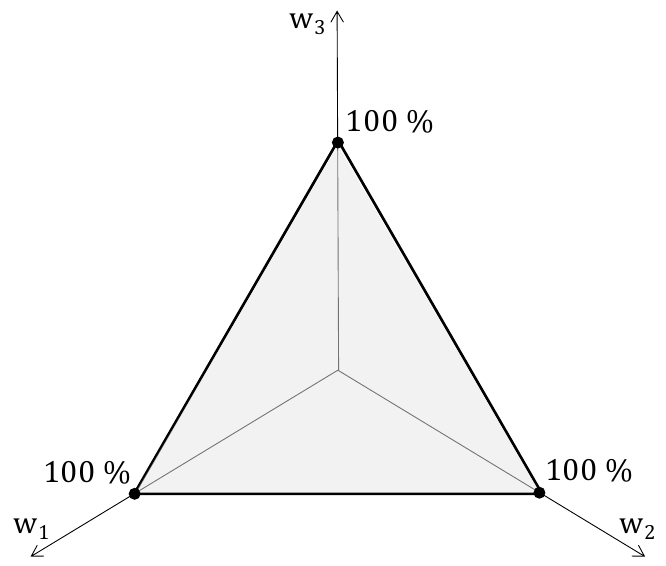}
 		\caption{\small Relative weight distributions 
 			among three voters in the unit simplex}
 		\label{fig:simplex}
 	\end{center}
 \end{figure}

Alas, it is generally an arduous task to determine for a given weight distribution $\mathbf{w}$ to which scoring equivalence class it belongs (for fixed vector $\mathbf{s}$).
The respective equivalence classes form convex polyhedra that are defined by linear inequalities. 
When we consider three players and restrict attention to their relative voting weights  $\mathbf{\bar w}=\mathbf{w}/(w_1+w_2+w_3)$ (so that $\bar w_1+\bar w_2+\bar w_3=100\%$), the polyhedra are either points, lines, or area pieces bounded by lines. They jointly cover the triangle highlighted in Figure~\ref{fig:simplex}~-- the so-called 2-dimensional unit simplex. 

Suppose that we have a `map' of all equivalence classes in this simplex. 
Then one may start out with an arbitrary weight distribution $\mathbf{w}=(w_1, w_2, w_3)$, compute the corresponding relative weight distribution $\mathbf{\bar w}$, locate it in the simplex map,
and now identify the applicable class. 

Such simplex maps can indeed be constructed. 
Namely, the figures depicted in Section~\ref{sec:Bilder} show the links from all possible weights to equivalence classes, except that we leave out a legend that would identify the respective equivalence classes via reference distributions of weights.\footnote{See Figure~5 in \citeN{Mayer/Napel:2021:Scoring}. It provides a map of the 51 Borda equivalence classes for $n=m=3$ and $w_1\ge w_2\ge w_3$ with a reference distribution of weights for each class. 
Maps could be constructed for more than three alternatives, too,
but the higher number of preference profiles and perturbations has considerable computational costs.
Equivalence classes for scores $0<s<1$ change fast: the number of Borda classes rises from 51 to 505 and $\ge\! 2251$ for $m=3$, $4$ and $5$ \shortcite{Kurz/Mayer/Napel:2019:WCG}. 
Corresponding analogues of Figure~\ref{fig:Die-Bilder} exhibit smoother transitions with even more shades of color.
} 

\section{Voting power and color}\label{sec:power_and_color}

Before we present our figures, let us explain how the selected coloring relates to the a~priori voting power of the three players involved. 
An index of voting power generally takes a description of a voting body~-- a simple voting game or, in our case, an $s$-scoring committee of three players deciding on three options~-- as its input and produces a real number for each player as its output.
The respective numbers reflect the players' influence on collective decisions according to a specific conception of a player being influential.
They are based on specific probabilistic assumptions about the voting situations faced by the players.

The popular Penrose-Banzhaf power index (\citeNP{Penrose:1946}; \citeNP{Banzhaf:1965}) equates `being influential' with the possibility of the considered player changing or swinging a decision at hand if the preferences and behavior of all other players are held fixed. 
This possibility arises, for instance, in unweighted binary majority decisions when the other players are split equally into a \emph{yes}-camp and a \emph{no}-camp, so that the player in question can determine which option receives the majority.
In other words, the considered player's vote is pivotal for the outcome.
The Penrose-Banzhaf index assesses the probability of pivotality events for a given player under the assumption that all other players vote \emph{yes} or \emph{no} with equal probabilities and independently of another. 
This is equivalent to assuming that all \emph{yes}-or-\emph{no} configurations or all coalitions $S\subseteq N$ of players who support a change of the status quo are equally likely.\footnote{
The Shapley-Shubik index \cite{Shapley/Shubik:1954} belongs to the same family of indices but supposes positive correlation of \emph{yes}-or-\emph{no} preferences across voters. 
In technical terms, it assumes an \emph{impartial anonymous culture (IAC)}, while the Penrose-Banzhaf index reflects an \emph{impartial culture (IC)}.\label{fn:IC_IAC}  
The Holler-Packel index \cite{Holler/Packel:1983} does not consider all coalitions of \emph{yes}-supporters but only \emph{minimal winning coalitions} $S\subseteq N$ in which every \emph{yes}-vote is pivotal for the outcome.
}

When the collective decision requires a choice from three or more options, such as candidates Ann, Bob and Clara above, one can similarly identify `being influential' with the committee's decision depending on or varying in the considered player's preferences.
For instance, if in our sports club example, player~3 did not rank Ann before Bob and Clara but had Bob as its first preference before Ann and Clara, then the plurality winner would be Bob rather than Ann.
Hence, player~3 is pivotal in the considered voting situation. 
So is player~2, whereas players~1 and 4 have no scope to individually change the winner for the given preferences of the respective others.
Players~1 and 4 are, however, pivotal for many other preference configurations $\mathbf{P}=(P_1,P_2,P_3,P_4)\in \mathcal{P}(A)^4$ that may arise.
So also they are influential from an a~priori perspective that considers all preference combinations to be possible.

Just as the Penrose-Banzhaf index is based on independent and equiprobable \emph{yes}-or-\emph{no} preferences in the binary case, players' preferences $P_i$ will be assumed to be distributed independently from the others also for more than two options, assigning equal probability to each of the conceivable strict orderings of the options. 
When assessing the a~priori influence implications of voting weights $\mathbf{w}=(5,4,3,1)$, for instance, we will therefore assume player~1 to be as likely to rank (i) Ann before Bob before Clara, as to rank (ii) Ann before Clara before Bob, (iii) Bob before Ann before Clara, (iv) Bob before Clara before Ann, (v) Clara before Ann before Bob, or (vi) Clara before Bob before Ann.
We allow the same six possibilities to arise independently also for players~2, 3 and 4.
So there are a total of $6\cdot 6\cdot 6 \cdot 6 =6^4 = 1296$ different voting situations that are equally probable when four players decide on three options.

We will focus here on only $n=3$ players who decide on $m=3$ options, so that $(m!)^n=6^3=216$ different preference profiles $\mathbf{P}=(P_1,P_2,P_3)$ are possible.
Holding a particular player of interest, say player~$i\in \{1,2,3\}$, fixed, we check for each profile whether a change of $i$'s ranking $P_i$ to one of the alternative five rankings $P_i'$ would make a difference to the collective decision.
Whenever this is the case, i.e., the profile $\mathbf{P'}$ that is created by replacing $P_i$ in $\mathbf{P}$ by $P_i'$ yields a collective decision $r^s|\mathbf{w}(\mathbf{P'})\neq r^s|\mathbf{w}(\mathbf{P})$, we count this as a \emph{swing position} for player~$i$.
Player~$i$'s power index value is then taken to be the ratio of actual swing positions to the maximum conceivable number of such positions. 

The latter corresponds to the number of swing positions that a dictator player would hold. 
For each of the 216 possible preference profiles of three voters on three options, the collective choice under a dictatorship equals the dictator's most preferred alternative.
So starting from given preferences of the dictator over three candidate, say ranking~(i) above, a switch to four of the five alternative rankings produces a different winner~-- namely preference changes from (i) to (iii), (iv), (v) or (vi). 
These perturbations involve a different top preference than (i) and let Bob or Claire win instead of Ann. 
It follows that a dictator player has $216\cdot 4=864$ swing positions:
they derive from considering 216 distinct voting situations and, for each situation, checking all five ways to spontaneously change the dictator's ranking of the options.
Such a change might reflect an idiosyncratic change of mind, perhaps due to new private information on the candidates; it might arise because the player is corrupt and sells its vote to an outside agent; it could simply be a demonstration of the player's power; etc.
If a player~$i$ in the actual scoring committee should have 432 swing positions, then the corresponding ratio $432/864=\sfrac{1}{2}$ reveals $i$ to be half as powerful as a dictator would be.

Expressing this reasoning in general mathematical terms leads to the (generalized) Penrose-Banzhaf index
\begin{equation}\label{eq:I_definition}
	\mathcal{PBI}_i(N,A,\rho)
	= \frac{\sum_{\mathbf{P} \in \mathcal{P}(A)^n} \sum_{P_i'\neq P_i \in \mathcal{P}(A)} \Delta \rho(\mathbf{P};{P'_i})}{m!^n\cdot (m!-(m-1)!)} 
\end{equation}
of {player $i$'s a priori influence} or \emph{voting power} in committee $(N,A,\rho)$, as introduced and axiomatically characterized by \shortciteN{Kurz/Mayer/Napel:2021:Influence}.\footnote{
Replacing the IC assumption that underlies eq.~\eqref{eq:I_definition} by the IAC assumption (cf.\ fn.~\ref{fn:IC_IAC}) naturally generalizes the Shapley-Shubik index (see \shortciteNP{Kurz/Mayer/Napel:2021:Influence}).
By contrast, generalization of the Holler-Packel index would first require the definition of a suitable analogue of minimal winning coalitions in weighted committee games. 
One possibility would be to study each winning alternative $a\in A$ separately and to consider \emph{$a$-minimal preference profiles} $\mathbf{P}$ where $\rho(\mathbf{P})=a$ such that 
$\rho(\mathbf{P}')\neq a$ for any profile $\mathbf{P}'$ in which $a$ is ranked lower by some voter with constant preferences on subset $A\smallsetminus a$.
}
Here $\Delta \rho(\mathbf{P};{P'_i})$ denotes an indicator function that is 1 if  $\rho(\mathbf{P'})\neq \rho(\mathbf{P})$, and 0 otherwise.
Equation~\eqref{eq:I_definition} is summing over all voting situations (i.e., all conceivable preference configurations $\mathbf{P}$), counts the number of changes of mind by player~$i$ (i.e., perturbations of $i$'s preferences $P_i$ to some $P_i'\neq P_i$) that change the collective decision, and then divides this by the total number of swing positions for a hypothetical dictator player ($6^3\cdot 4=864$ for $m=3$ options and $n=3$ players).
So for an $s$-scoring committee $(N,A,r^s|\mathbf{w})$ of three players, the triplet 
$$
\big(\mathcal{PBI}_1(N,A,r^s|\mathbf{w}),\mathcal{PBI}_2(N,A,r^s|\mathbf{w}),\mathcal{PBI}_3(N,A,r^s|\mathbf{w})\big),
$$ 
or $\mathcal{PBI}$ for short, quantifies the distribution of voting power in the committee in terms of how close the individual players are to having dictatorial influence. In our sports club example, the power distribution amounts to $\mathcal{PBI}\approx(0.6296, 0.4815, 0.4444, 0.0741)$. That is, player 1 has about 63\% of the influence of a dictator while player 4 only has about 7\% of the influence of a dictator. The influence of players two and three is just under 50\% of that of a dictator. 

For graphical purposes one might now associate player~1's power value $\mathcal{PBI}_1$ with the color red, player~2's power $\mathcal{PBI}_2$ with green and player~3's power $\mathcal{PBI}_3$ with blue.
Thus we would have linked the scoring rule $r^s$ for a given value of $s$ and a particular distribution $\mathbf{w}$ of voting weights to a particular color using the common RGB color code. 
For instance, $\mathcal{PBI}(N,A,r^s|(1,0,0))=(1,0,0)$ for any $0\le s\le 1$ and this would correspond to bright red color. 
Or the power distribution $\mathcal{PBI}=(\sfrac{588}{864}, \sfrac{516}{864},\sfrac{312}{864})\approx(0.6806, 0.5972, 0.3611)$ that is derived by \shortciteN{Kurz/Mayer/Napel:2021:Influence} for $s=\sfrac{1}{2}$ and weights $\mathbf{w}=(6,5,3)$ would correspond to a dark khaki color. 

Although this would be feasible, the figures in Section~\ref{sec:Bilder} will not use exactly this coloring option.
We will rather make two modifications:
first, we will adopt a structural view on committee equivalences, i.e., we do not consider player labels important. 
Hence we will give the same color to all six points in the unit simplex that represent relative voting weights of, e.g., $\mathbf{\bar w}=(\sfrac{6}{14},\sfrac{5}{14},\sfrac{3}{14})$ after sorting the weights in decreasing order.
This implies that the coloring of the weight simplex will be 3-fold radially symmetric around $\mathbf{\bar w}=(\sfrac{1}{3},\sfrac{1}{3},\sfrac{1}{3})$, as well as mirror symmetric with the three symmetry axes $\bar w_1=\bar w_2$, $\bar w_2=\bar w_3$ and $\bar w_1=\bar w_3$.

Second, we will apply a transformation when turning power triplets $\mathcal{PBI}$ into RGB levels.
The motivation is to make better use of the available color palette, to obtain a somewhat lighter image than by, e.g., associating $\mathbf{\bar w}$ with dark khaki, and to represent dictatorial power by the dark blue color that has already been used, e.g., by \shortciteN{Kurz/Mayer/Napel:2019:WCG}.


\section{Simplex maps of equivalence classes}\label{sec:Bilder}

All images displayed in this section are derived via the following five steps:
\begin{enumerate}
	\item We fix a scoring vector $\mathbf{s}=(1,s,0)$ and consider the corresponding scoring rule $r^s$ for collective decisions on $m=3$ options by $n=3$ players.
	\item We use a finite grid of rational numbers and let the computer loop through all relative voting weight distributions $\mathbf{\bar w}$ with $1\ge \bar w_1\ge \bar w_2\ge \bar w_3\ge 0$ on this grid.
	\item For each of the 282\,376 weight distributions $\mathbf{\bar w}^k$, $k=1,2,\dots,282\,376$, on the adopted grid, we compute the Penrose-Banzhaf voting power $\mathcal{PBI}^k$ in the respective weighted $s$-scoring committee $(N,A,r^s|\mathbf{\bar w}^k)$.
	\item The obtained triplet $(\mathcal{PBI}_1^k,\mathcal{PBI}_2^k,\mathcal{PBI}_3^k)$ is then transformed into red, green and blue intensities $(R,G,B)=\big(\frac{2\cdot\mathcal{PBI}_3^k}{\max_k \mathcal{PBI}_3^k}, \frac{\mathcal{PBI}_2^k}{\max_k \mathcal{PBI}_2^k}, \frac{\mathcal{PBI}_1^k-\min_k \mathcal{PBI}_1^k}{1-\min_k\mathcal{PBI}_1^k})$.
	\item For each weight distribution  $\mathbf{\bar w}$, the six points in the simplex (cf.\ Figure~\ref{fig:simplex}) that structurally correspond to  $\mathbf{\bar w}$~-- that is, $(\bar w_1, \bar w_2, \bar w_3 )$, $(\bar w_2, \bar w_1 ,\bar w_3 )$, $(\bar w_1, \bar w_3, \bar w_2 )$, etc.~-- are colored with the RGB intensities given by $(R,G,B)$. 
	For instance, $\mathcal{PBI}$ figures of $(1,0,0)$ translate into $(R,G,B)=(0,0,1)$ and dark blue color.
\end{enumerate}

It is noteworthy that the distribution of voting power in two $s$-scoring committees $(N,A,r^s|\mathbf{w})$ and  $(N,A,r^s|\mathbf{w'})$ can coincide even though the committees are non-equivalent: players are exactly as influential in either but some preference profiles yield different decisions so that $r^s|\mathbf{w}\not \equiv r^s|\mathbf{w'}$. 
Some of the illustrations in Figure~\ref{fig:Die-Bilder} therefore involve fewer different colors than there are distinct equivalence classes for the considered value of $s$. 
Moreover, equivalence classes that are represented by a single point in the simplex like the symmetric distribution of relative voting weights $\mathbf{\bar w}=(\sfrac{1}{3},\sfrac{1}{3}, \sfrac{1}{3})$, or a line~-- e.g., $\mathbf{\bar w}=(x,1-x,0)$ for $0<x<\sfrac{1}{2}$~-- may not be visible without magnification.
We have manually enlarged them only for $s=0$ and $s=1$.
Bearing these caveats in mind the colored simplices below provide accurate maps of all equivalence classes of scoring committees that exist for a given value of $s$.

\def\breite{0.490} 

\def\titelabstand{-0.44}
\def\panelabstand{0.41}

\newpage
\begin{figure}[h!]
	\centering
	\includegraphics[width=\breite\textwidth]{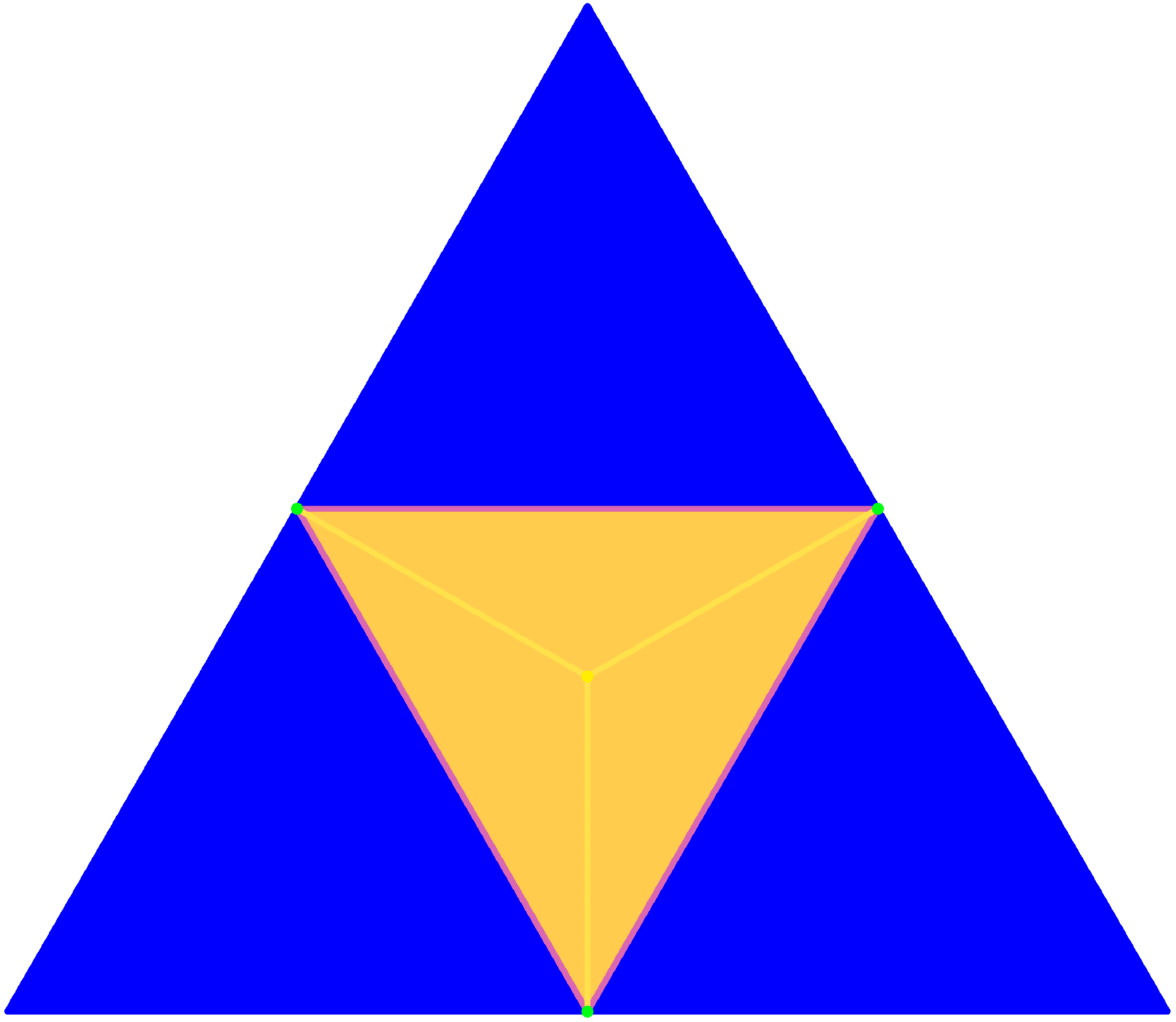}\enspace
	\includegraphics[width=\breite\textwidth]{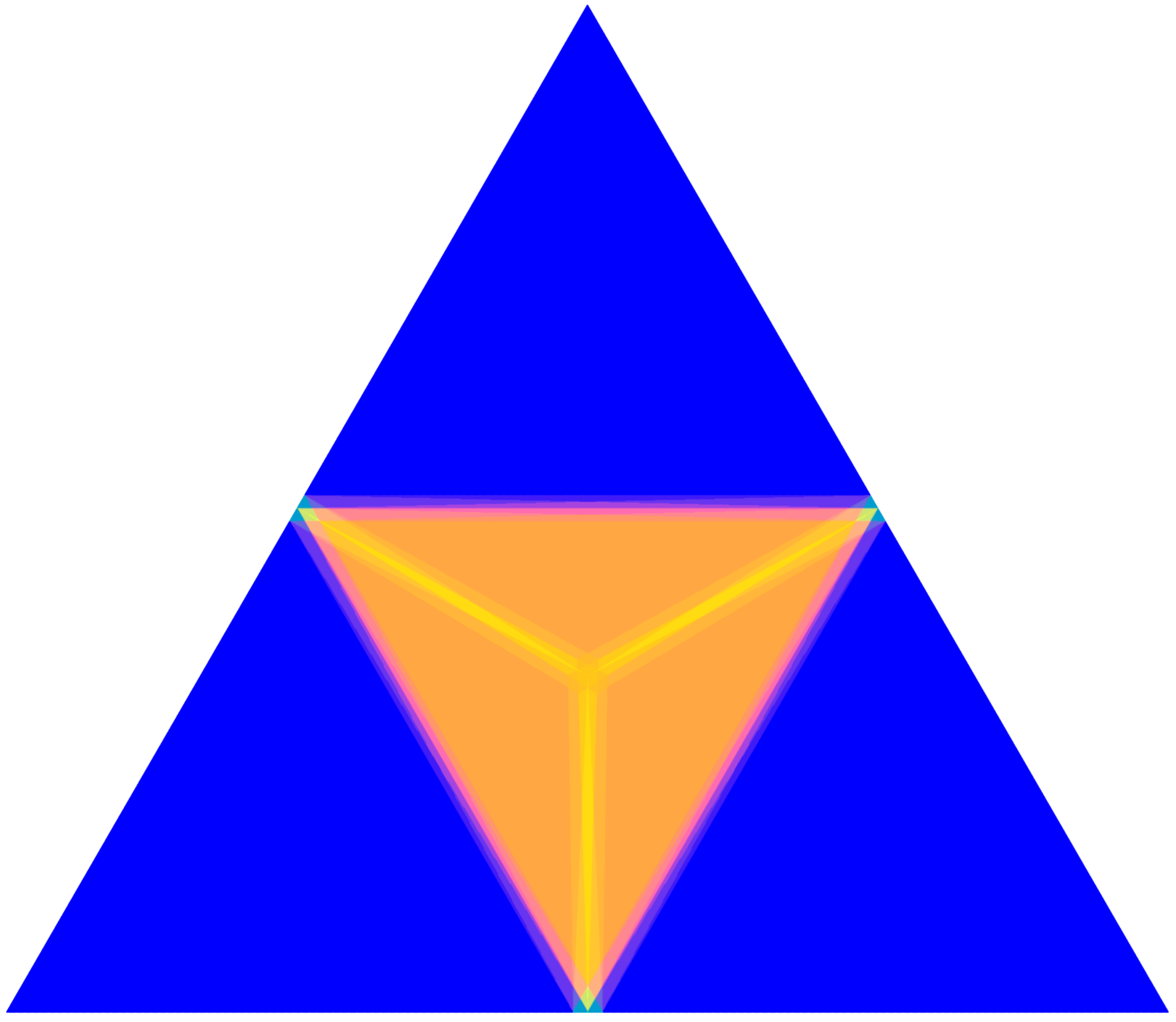}
	\vspace{\titelabstand\textwidth}
	\
	
	{\small
	\makebox[\breite\textwidth][l]{$s=0$ (plurality):}\enspace
	\makebox[\breite\textwidth][l]{$s=0.05$:}}
	
	\vspace{\panelabstand\textwidth}
	
\includegraphics[width=\breite\textwidth]{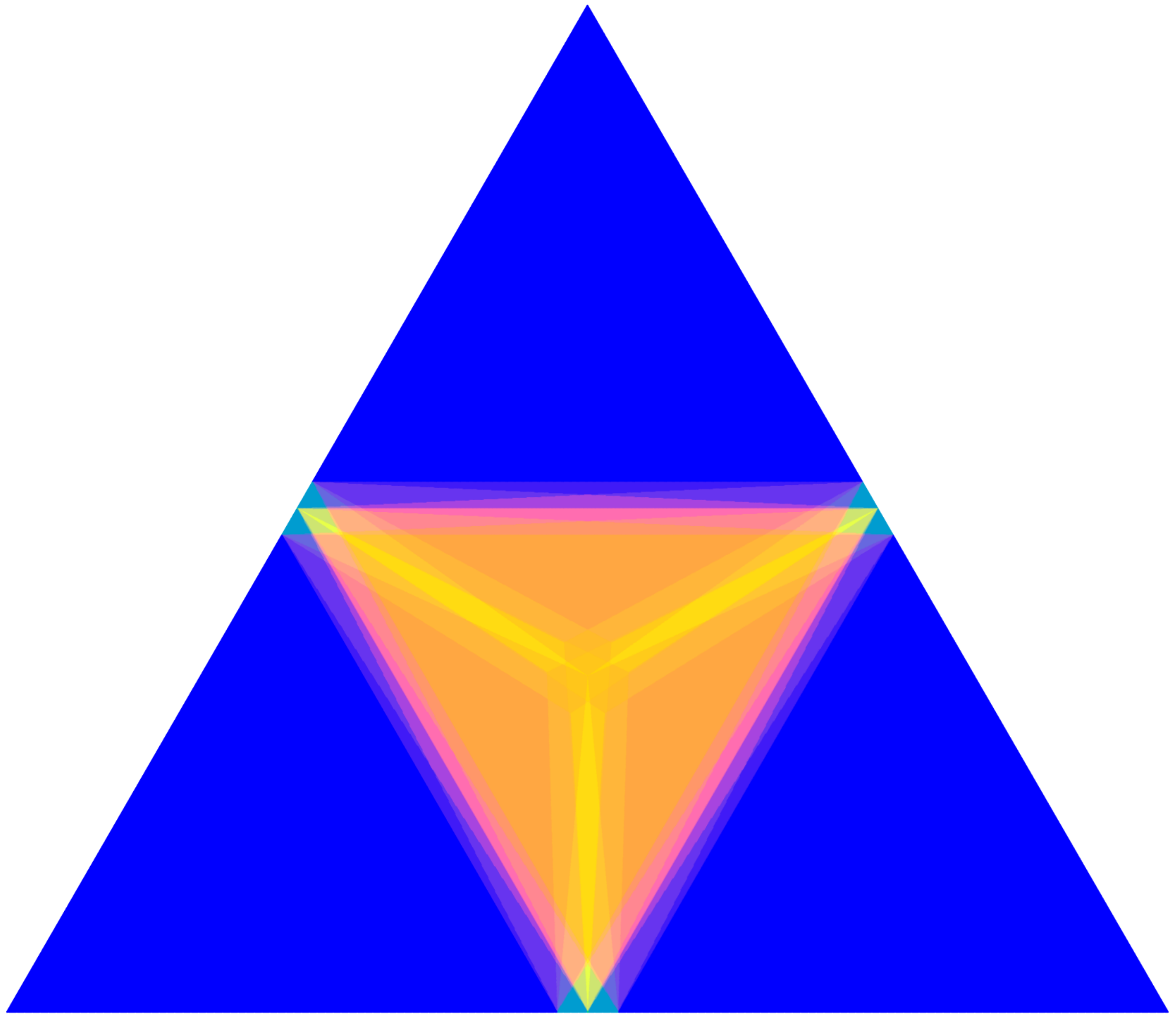}\enspace
\includegraphics[width=\breite\textwidth]{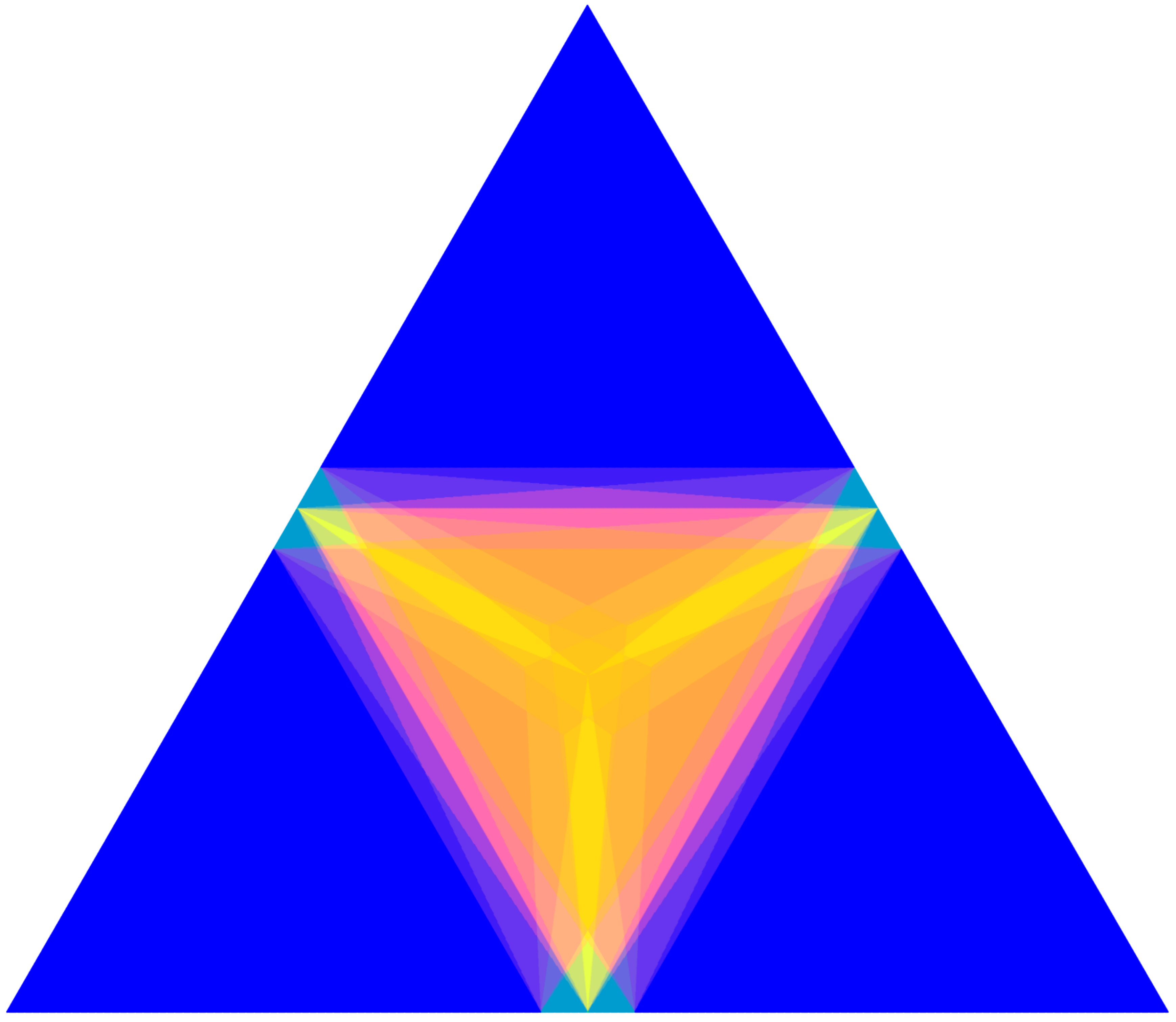}
	\vspace{\titelabstand\textwidth}
\

{\small
	\makebox[\breite\textwidth][l]{$s=0.10$:}\enspace
	\makebox[\breite\textwidth][l]{$s=0.15$:}}

	\vspace{\panelabstand\textwidth}
	
\includegraphics[width=\breite\textwidth]{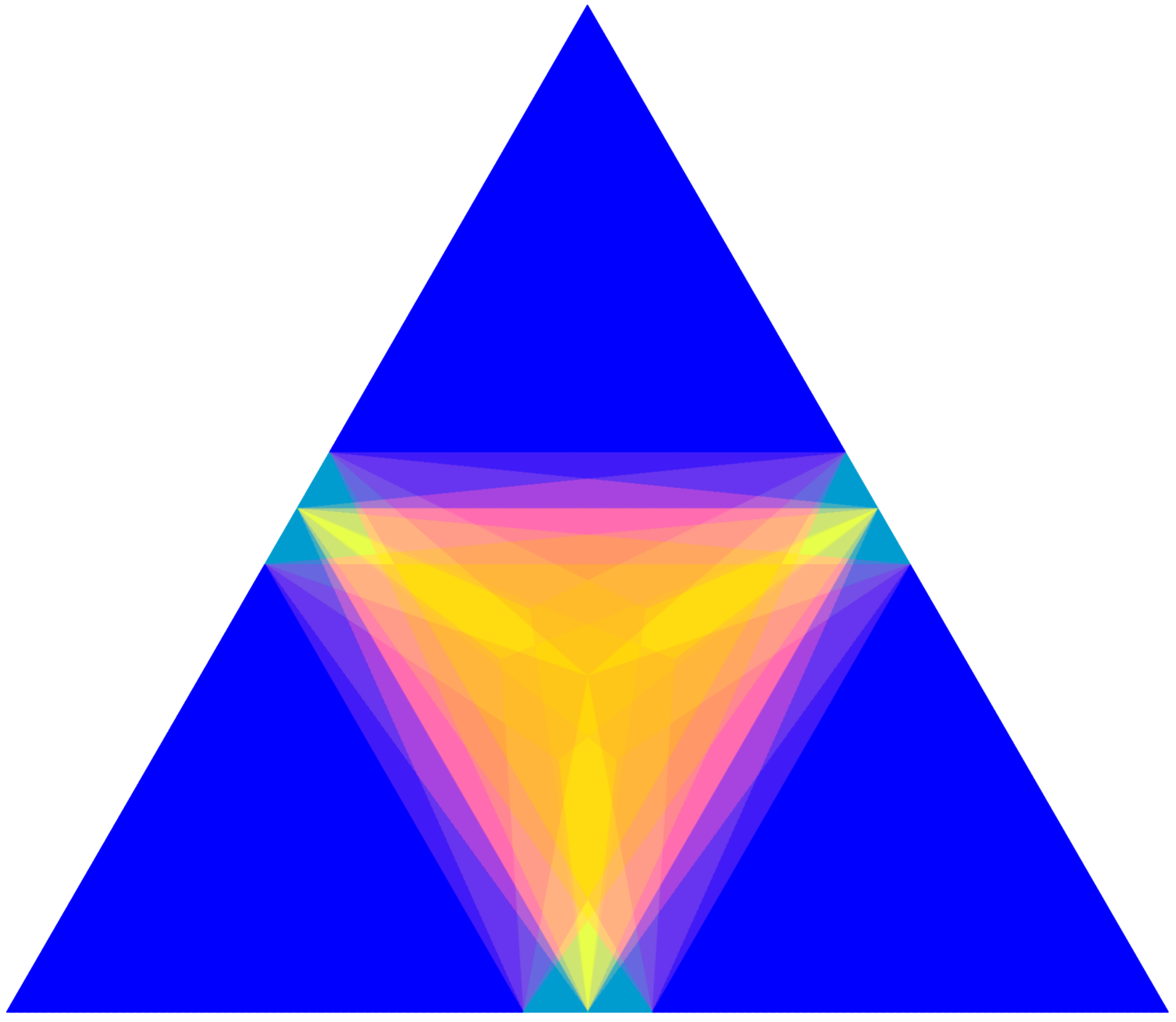}\enspace
\includegraphics[width=\breite\textwidth]{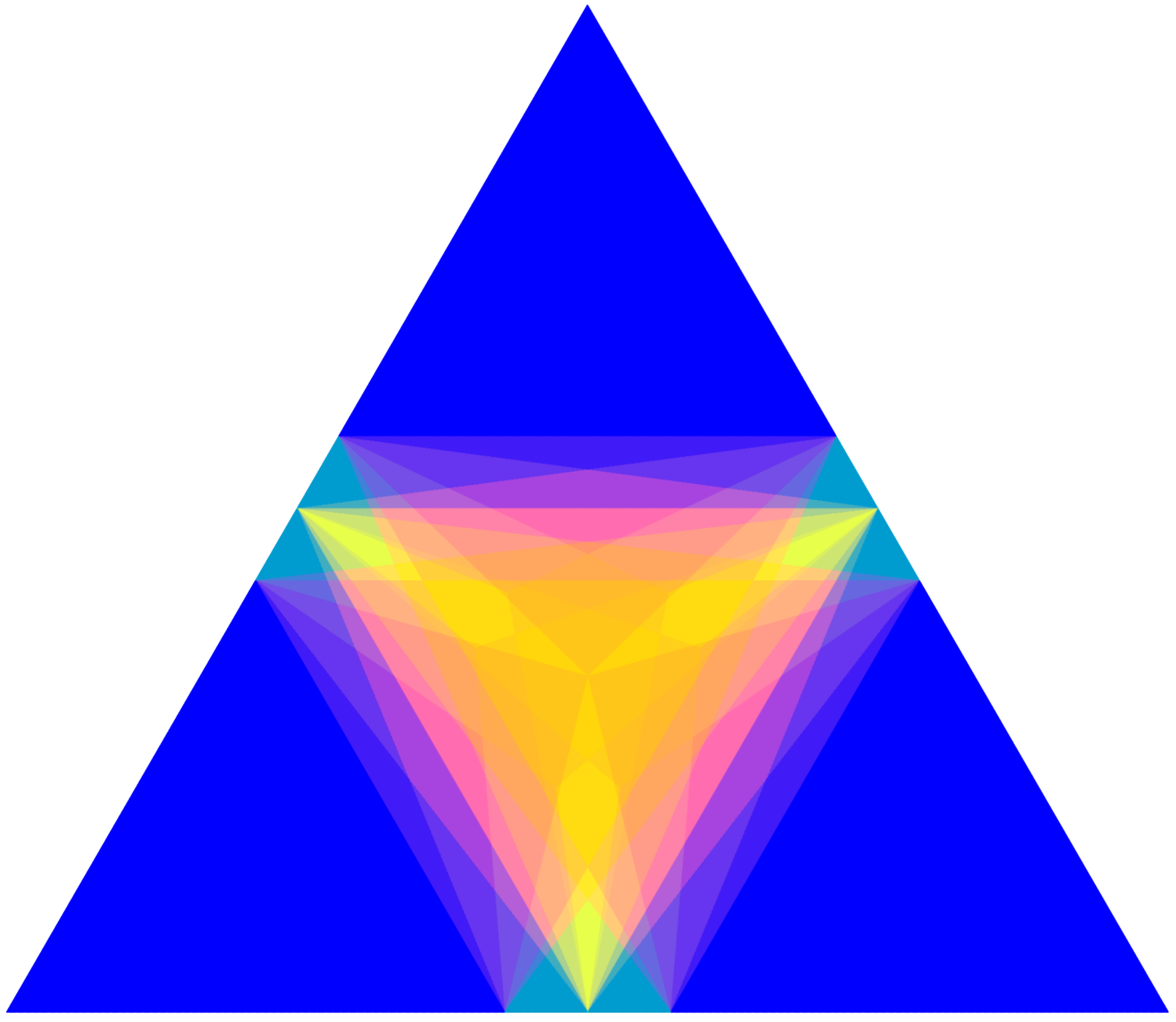}
	\vspace{\titelabstand\textwidth}
\

{\small
	\makebox[\breite\textwidth][l]{$s=0.20$:}%
	\makebox[\breite\textwidth][l]{$s=0.25$:}}

	\vspace{\panelabstand\textwidth}

	\caption{Weighted $s$-scoring committees in Penrose-Banzhaf coloring }
	\label{fig:Die-Bilder}
\end{figure}

\begin{figure}[h!]
	\centering
	\includegraphics[width=\breite\textwidth]{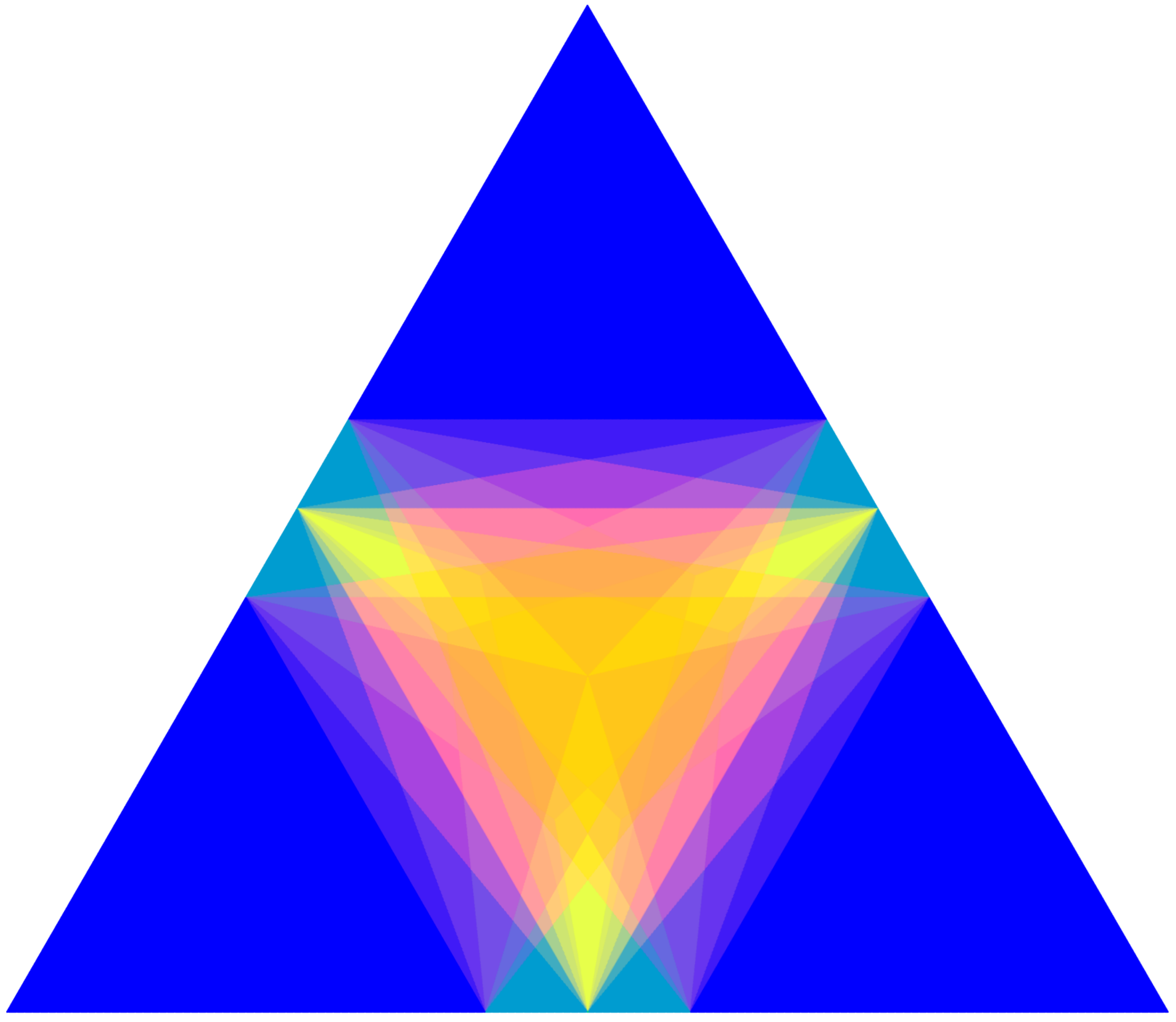}\enspace
	\includegraphics[width=\breite\textwidth]{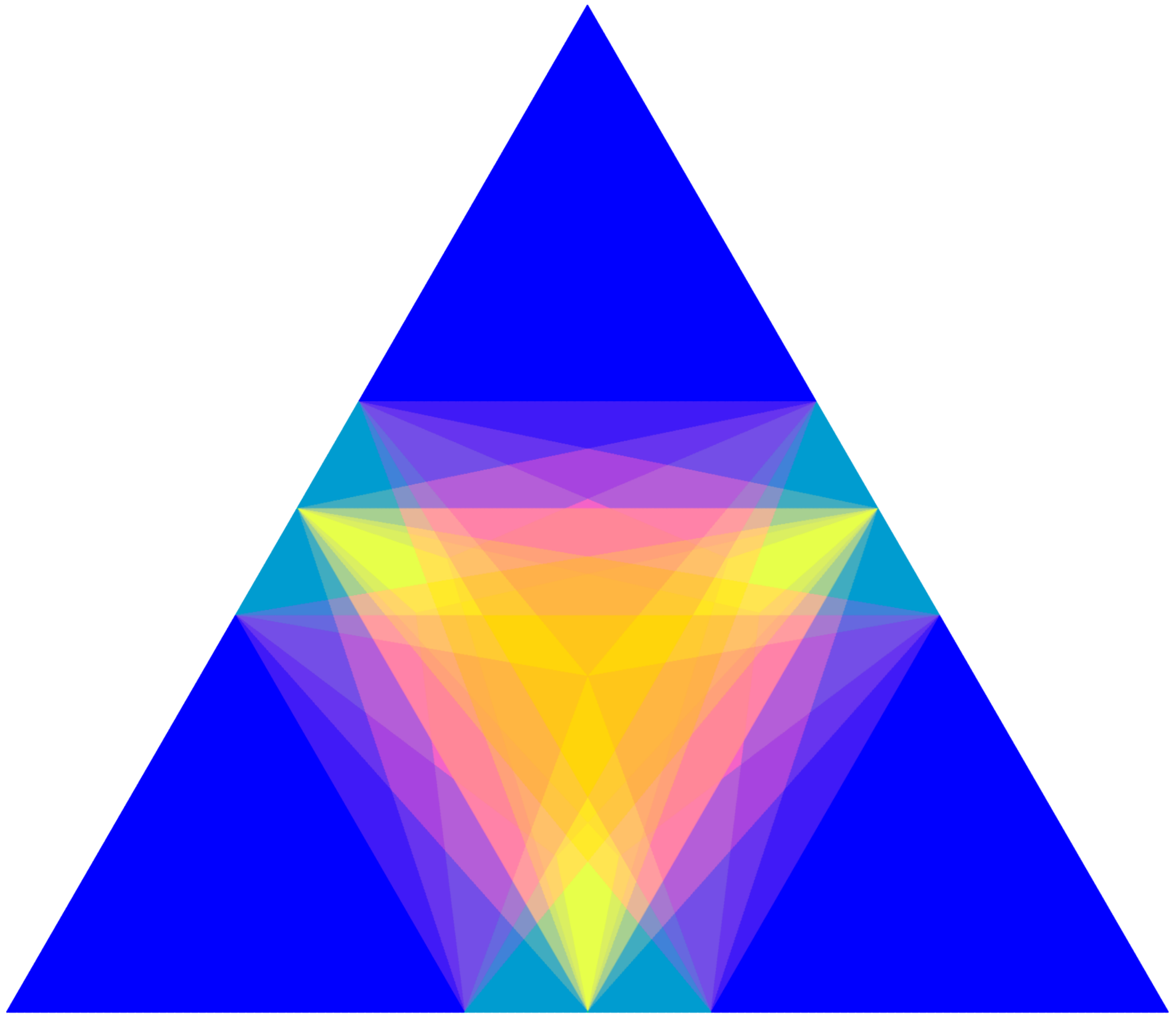}
	\vspace{\titelabstand\textwidth}
	\
	
	{\small
		\makebox[\breite\textwidth][l]{$s=0.30$:}\enspace
		\makebox[\breite\textwidth][l]{$s=0.35$:}}
	
	\vspace{\panelabstand\textwidth}
	
	\includegraphics[width=\breite\textwidth]{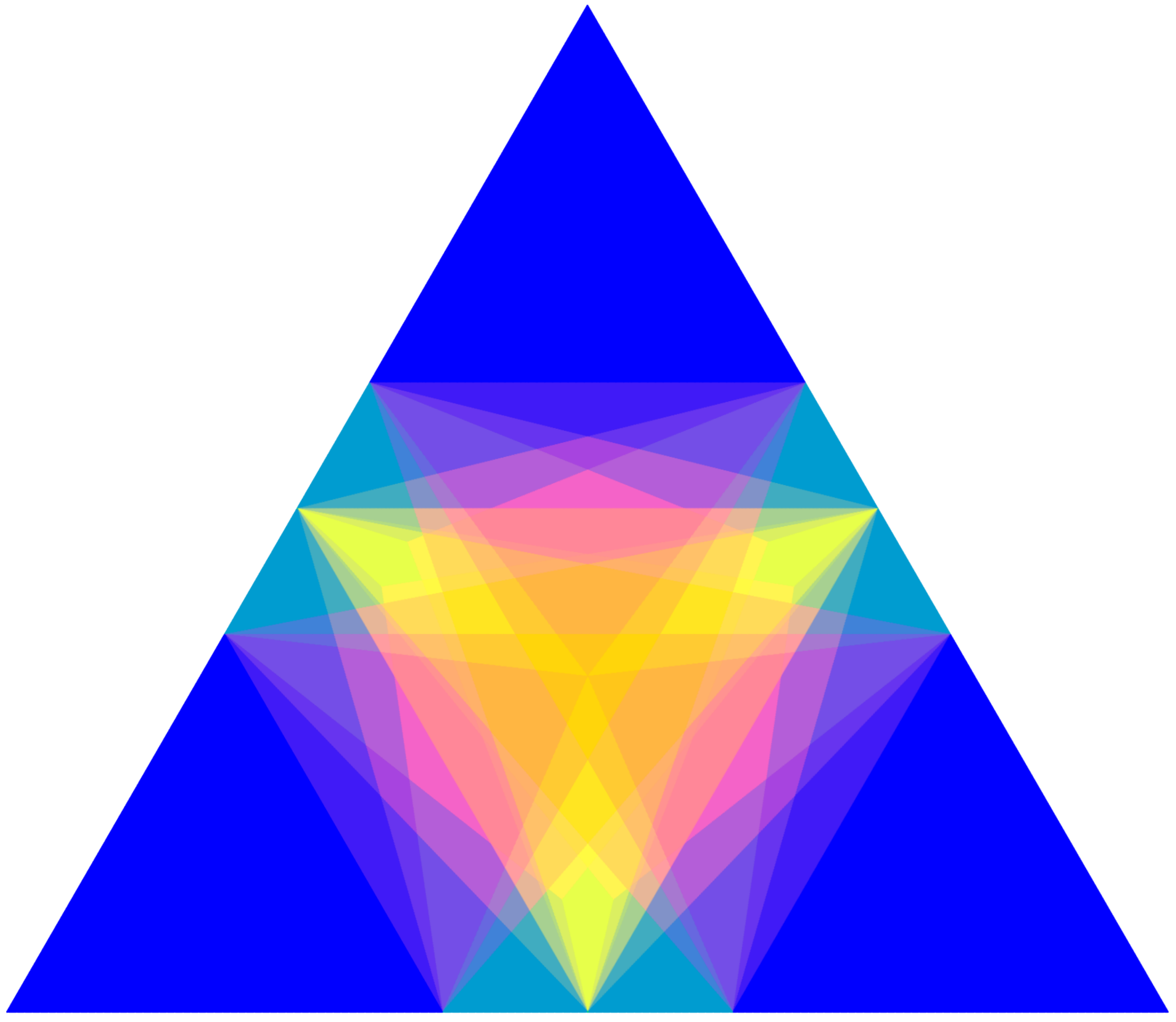}\enspace
	\includegraphics[width=\breite\textwidth]{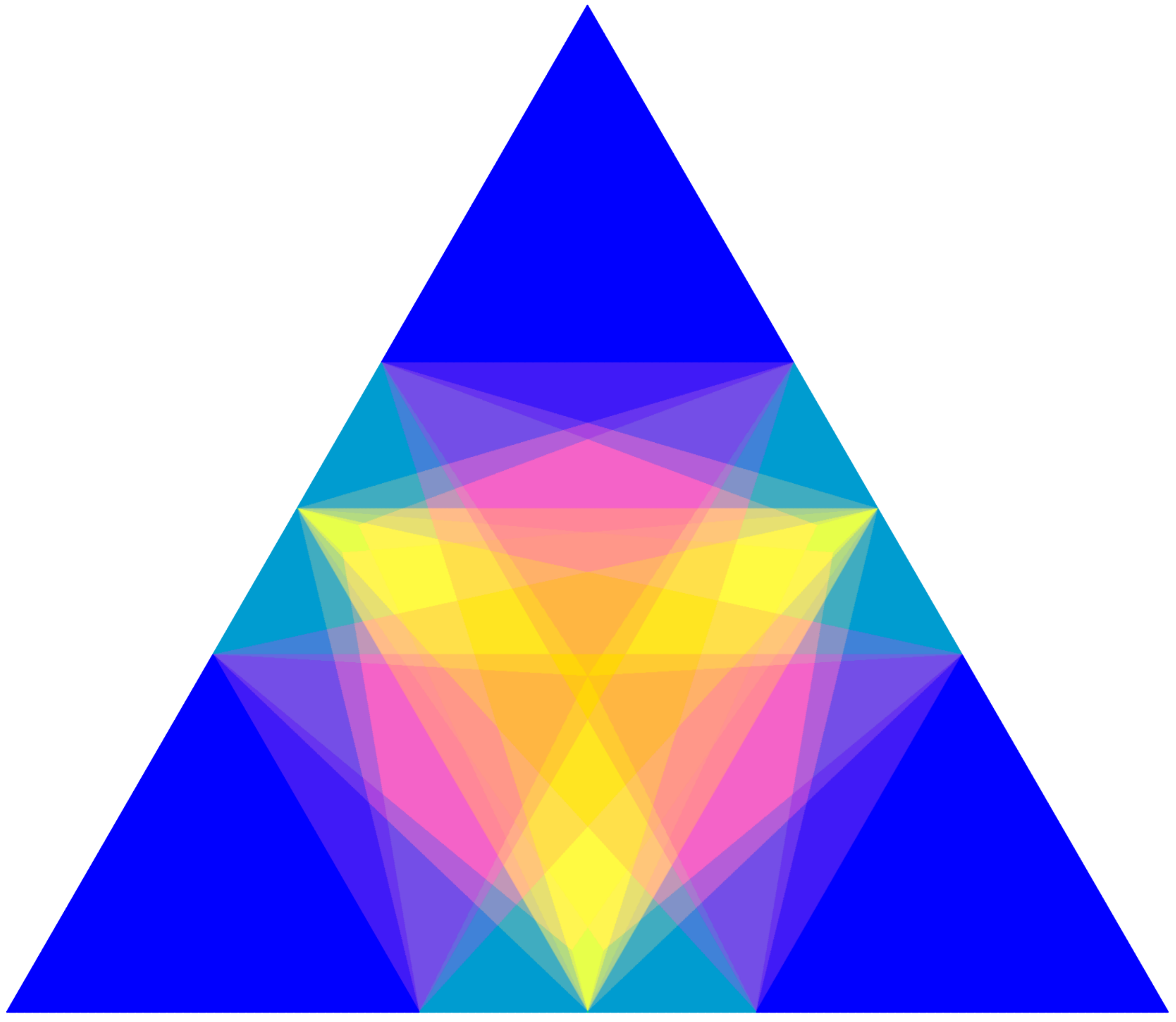}
	\vspace{\titelabstand\textwidth}
	\
	
	{\small
		\makebox[\breite\textwidth][l]{$s=0.40$:}\enspace
		\makebox[\breite\textwidth][l]{$s=0.45$:}}
	
	\vspace{\panelabstand\textwidth}
	
	\includegraphics[width=\breite\textwidth]{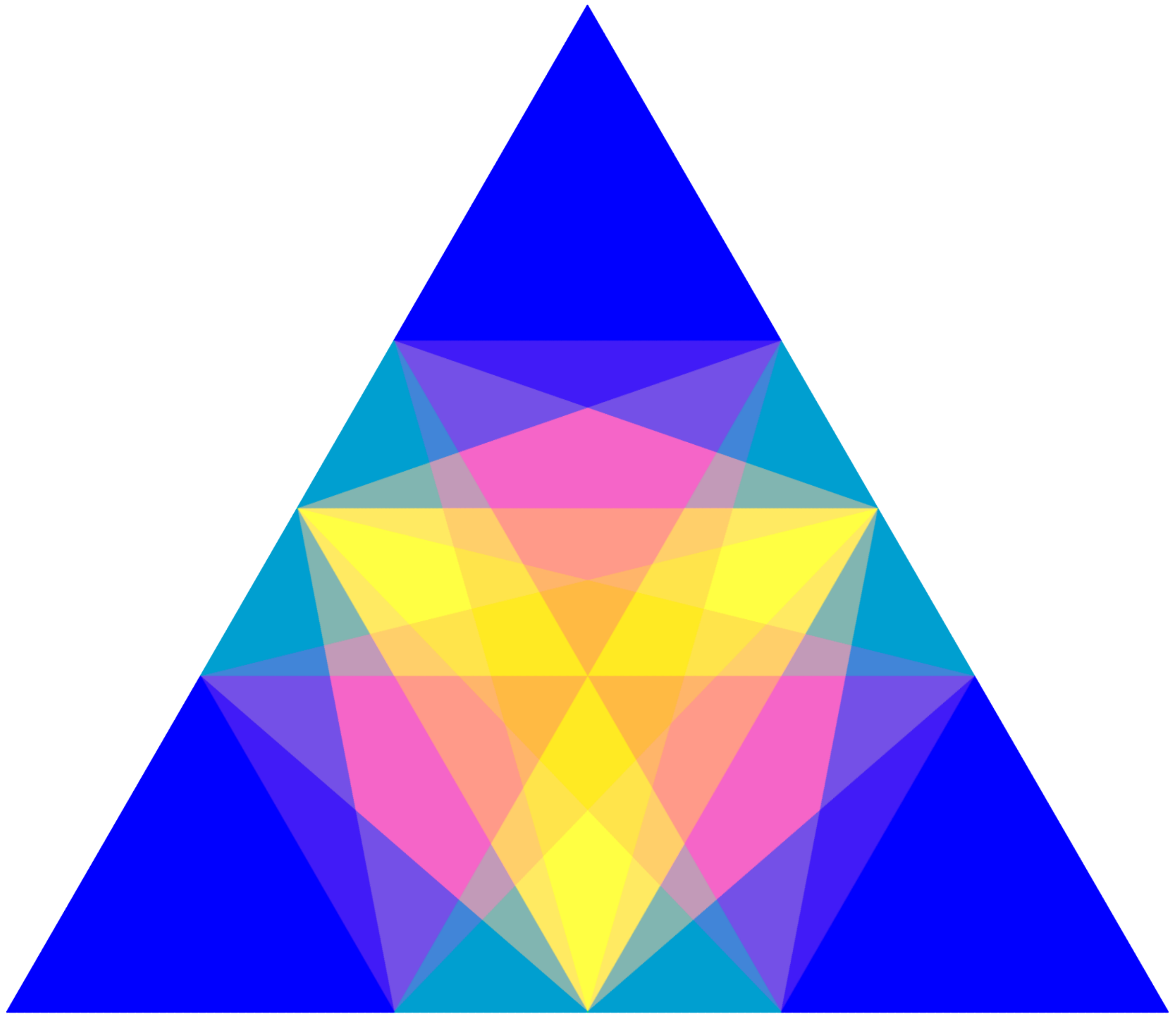}\enspace
	\includegraphics[width=\breite\textwidth]{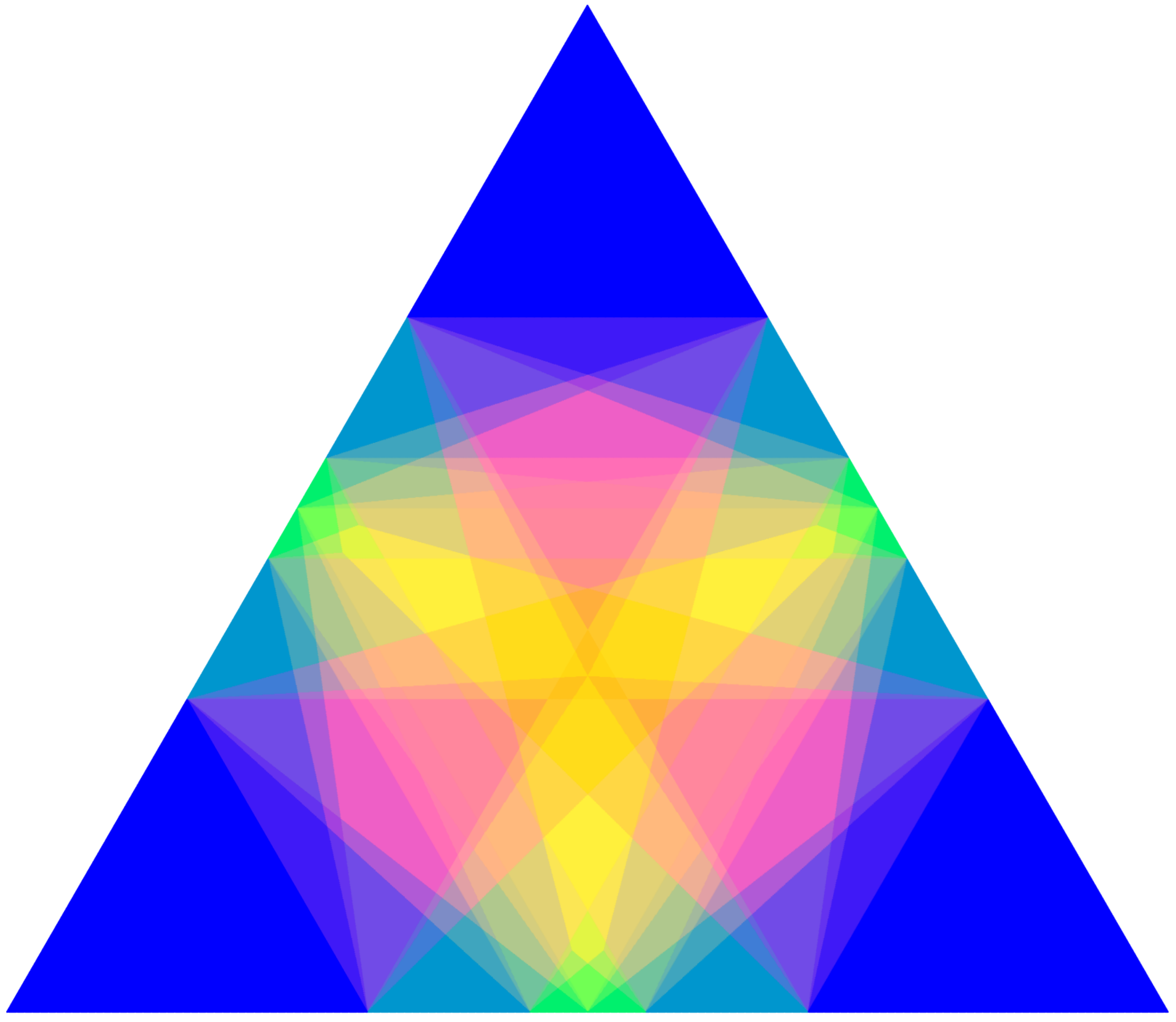}
	\vspace{\titelabstand\textwidth}
	\
	
	{\small
		\makebox[\breite\textwidth][l]{$s=0.5$ (Borda):}%
		\makebox[\breite\textwidth][l]{$s=0.55$:}}
	
	\vspace{\panelabstand\textwidth}
	
	\medskip
	{\captionsize Figure 3 (ctd.): Weighted $s$-scoring committees in Penrose-Banzhaf coloring }
\end{figure}

\begin{figure}[h!]
	\centering
	\includegraphics[width=\breite\textwidth]{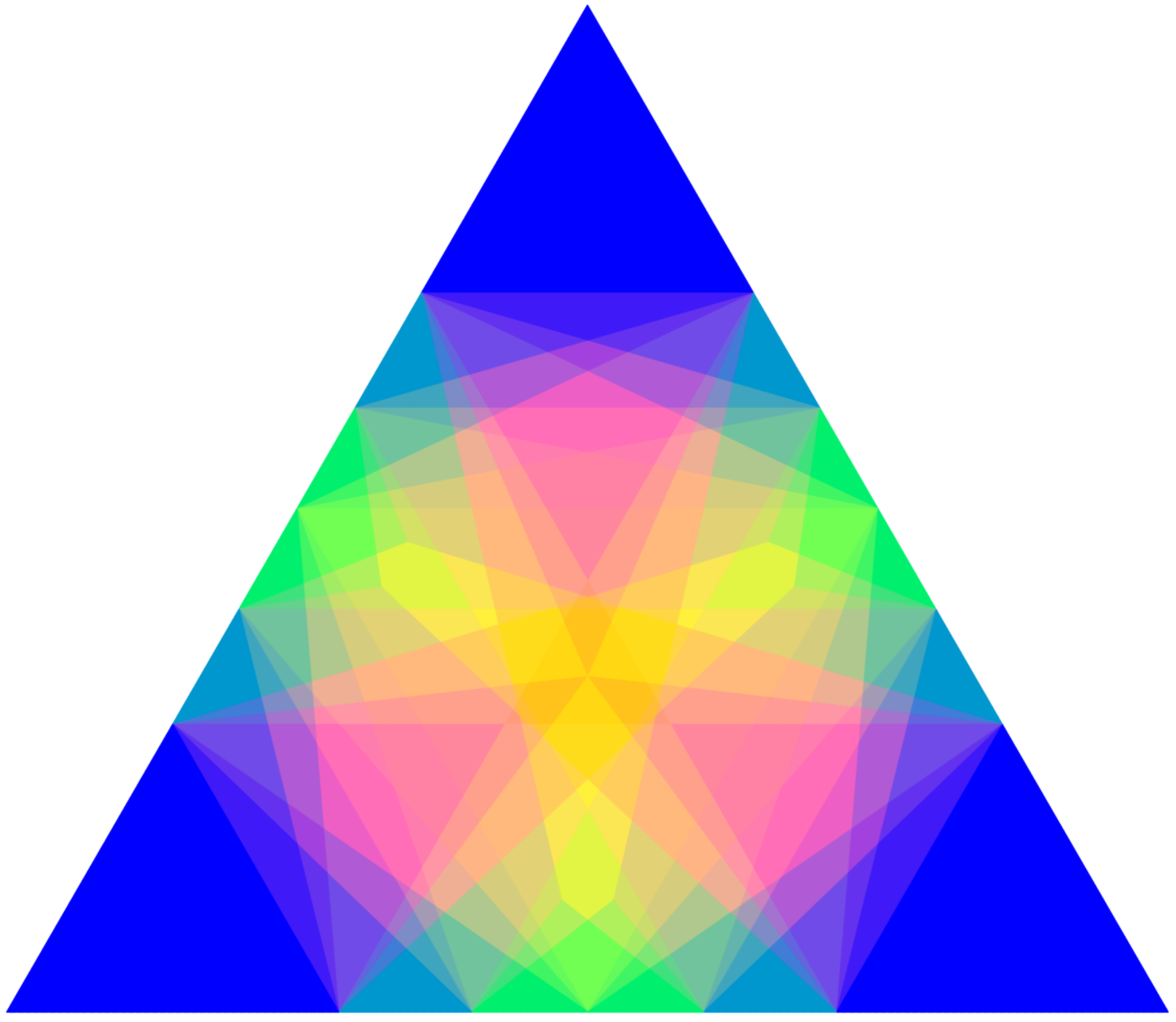}\enspace
	\includegraphics[width=\breite\textwidth]{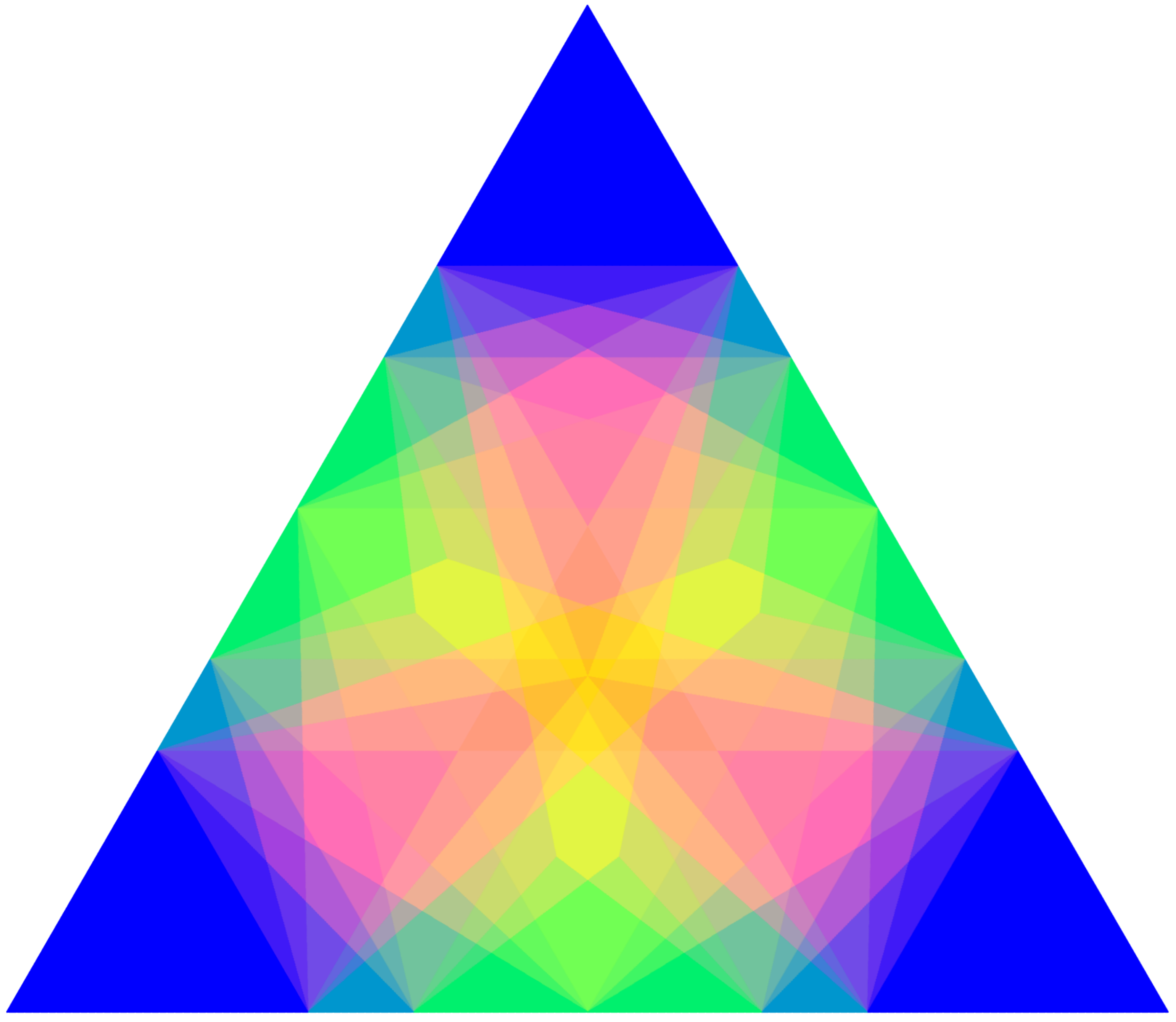}
	\vspace{\titelabstand\textwidth}
	\
	
	{\small
		\makebox[\breite\textwidth][l]{$s=0.60$:}\enspace
		\makebox[\breite\textwidth][l]{$s=0.65$:}}
	
	\vspace{\panelabstand\textwidth}
	
	\includegraphics[width=\breite\textwidth]{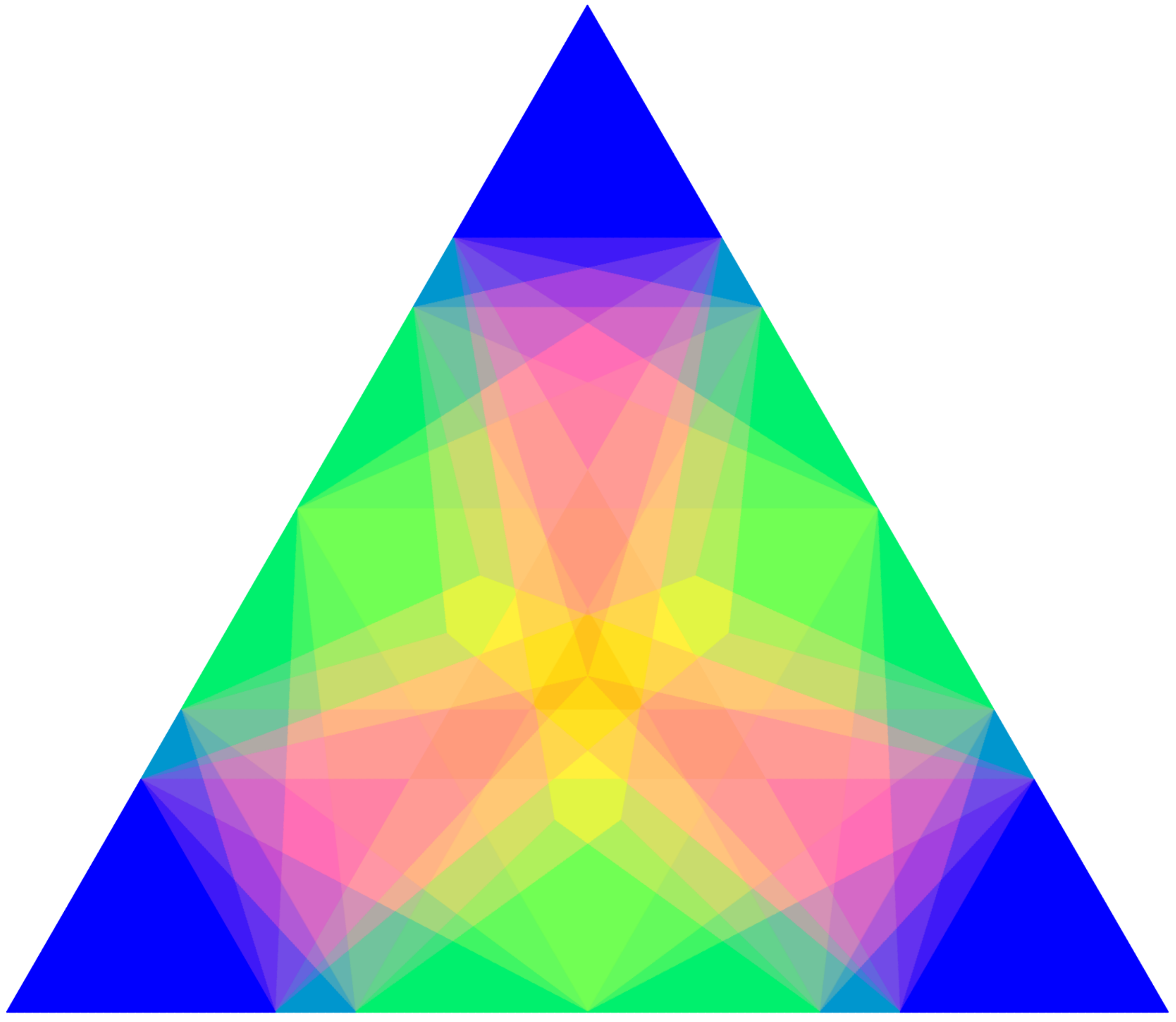}\enspace
	\includegraphics[width=\breite\textwidth]{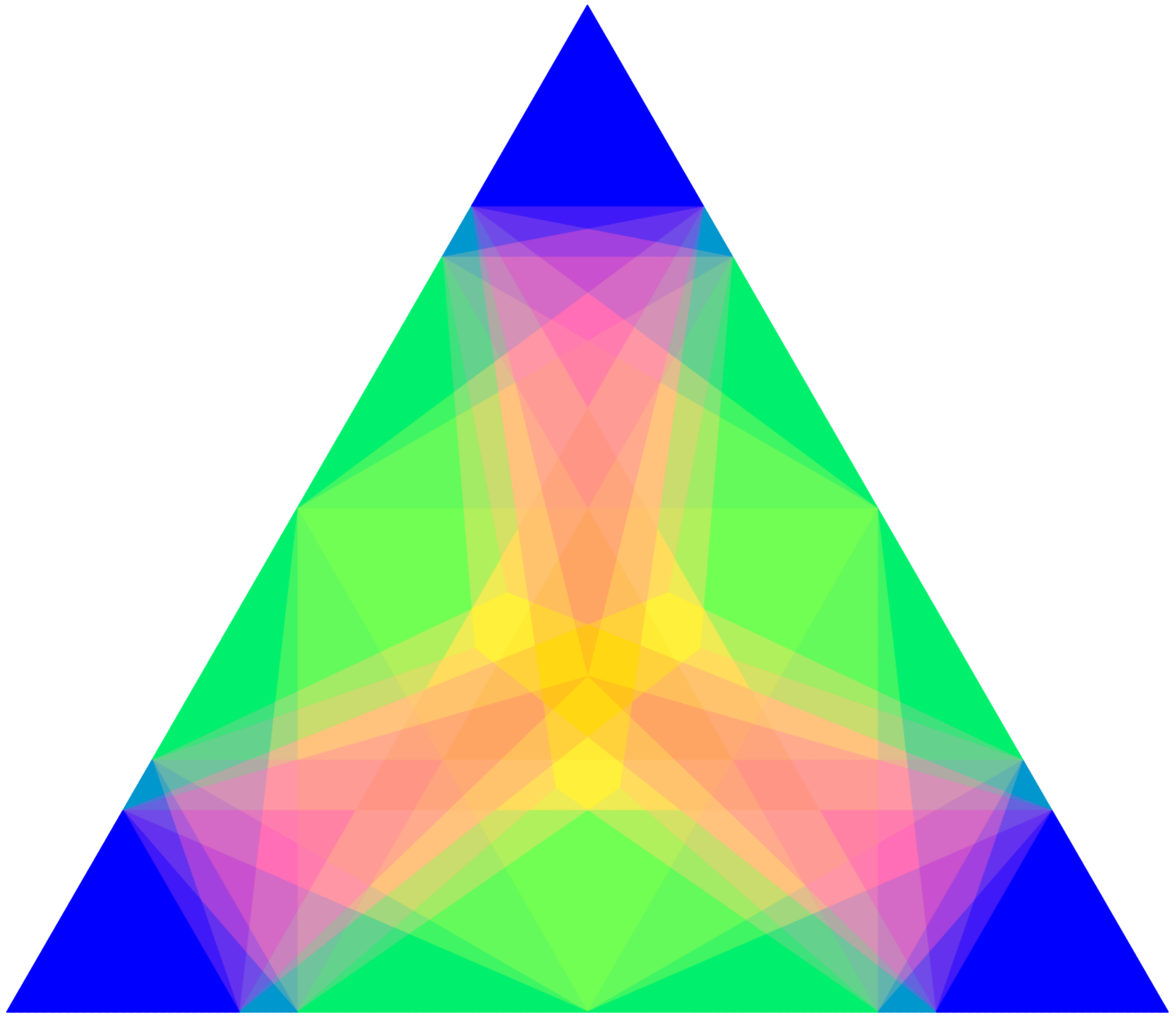}
	\vspace{\titelabstand\textwidth}
	\
	
	{\small
		\makebox[\breite\textwidth][l]{$s=0.70$:}\enspace
		\makebox[\breite\textwidth][l]{$s=0.75$:}}
	
	\vspace{\panelabstand\textwidth}
	
	\includegraphics[width=\breite\textwidth]{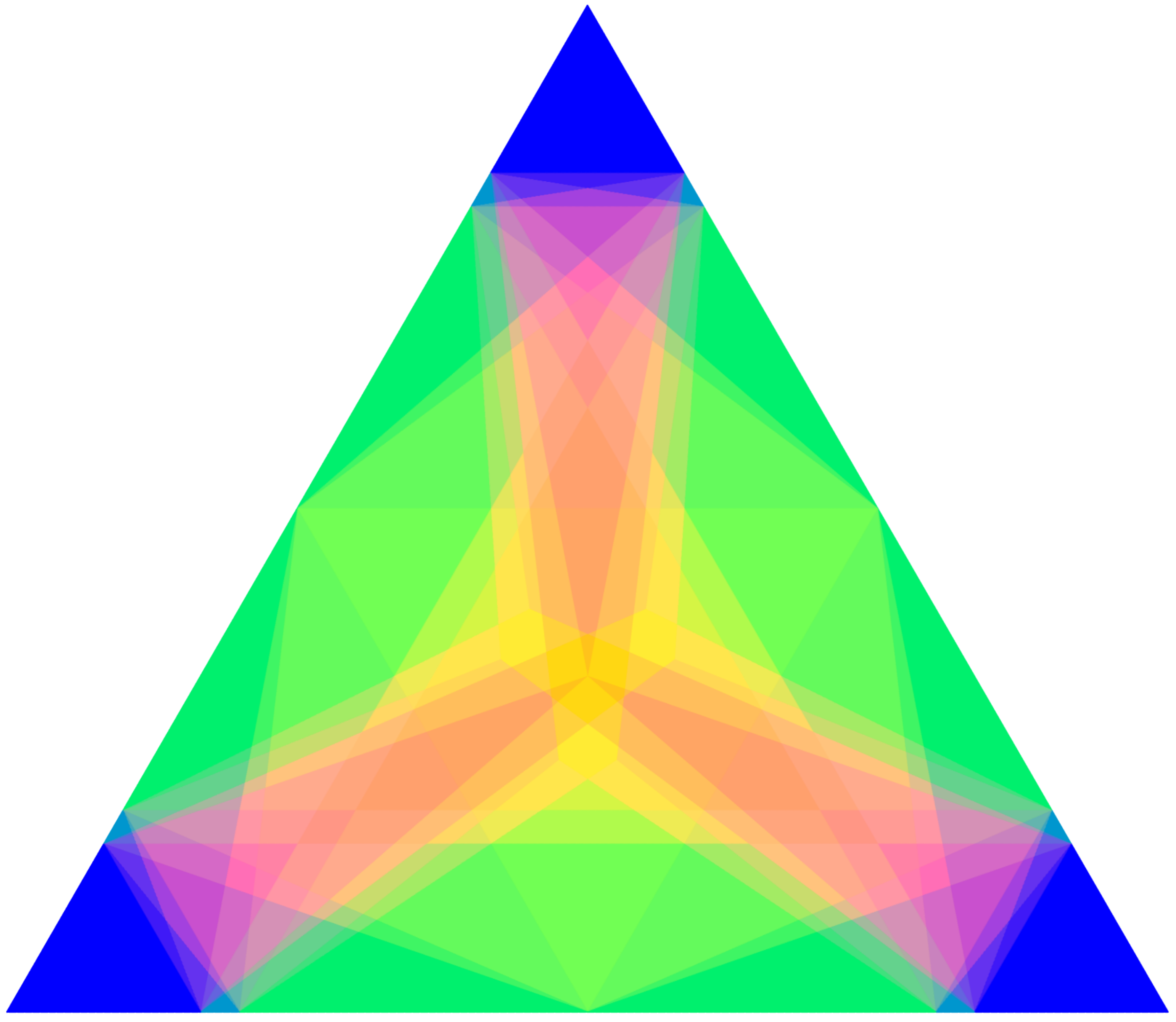}\enspace
	\includegraphics[width=\breite\textwidth]{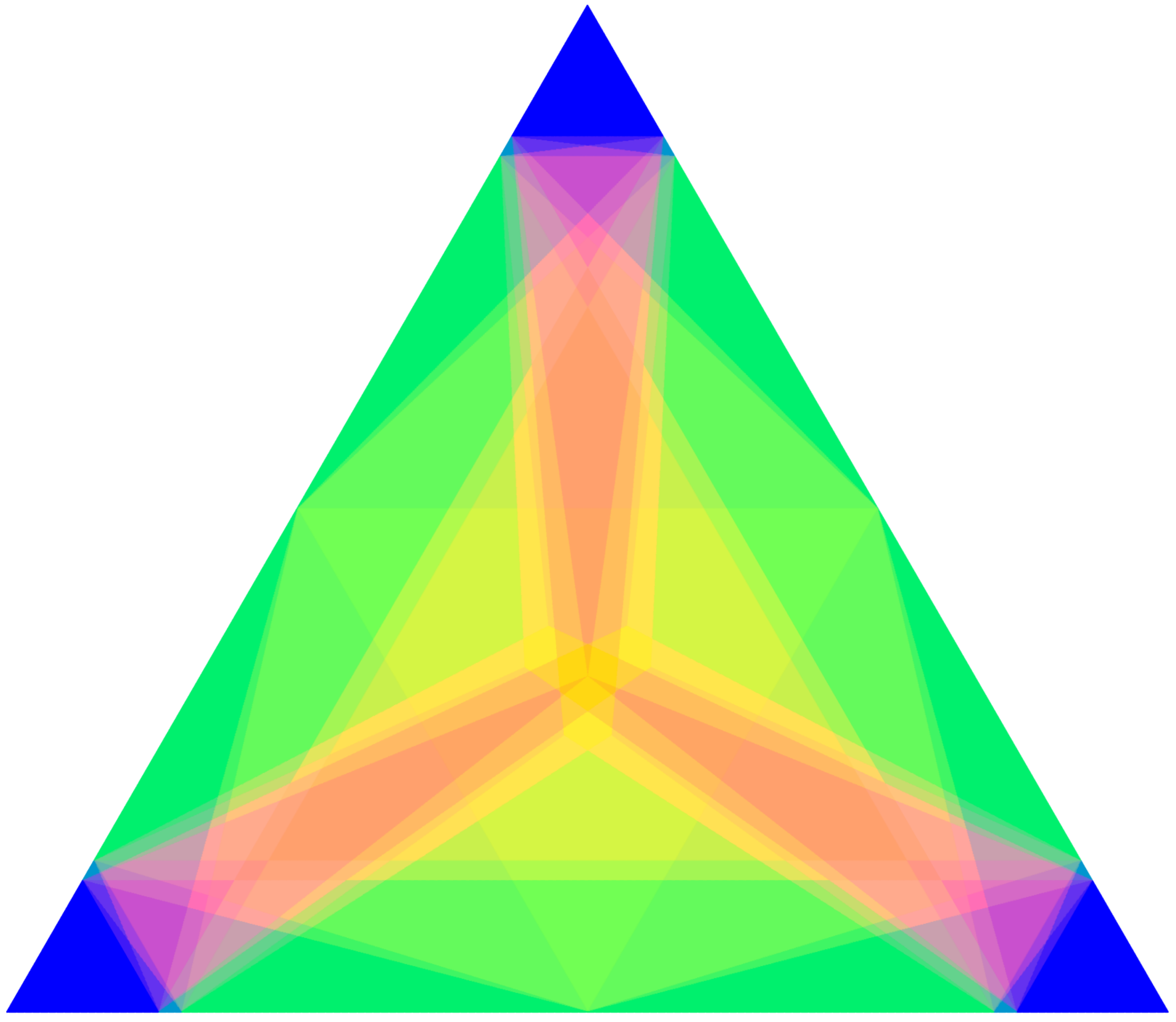}
	\vspace{\titelabstand\textwidth}
	\
	
	{\small
		\makebox[\breite\textwidth][l]{$s=0.80$:}%
		\makebox[\breite\textwidth][l]{$s=0.85$:}}
	
	\vspace{\panelabstand\textwidth}
	
	\medskip
	{\captionsize Figure 3 (ctd.): Weighted $s$-scoring committees in Penrose-Banzhaf coloring }
\end{figure}

\begin{figure}[h!]
	\centering
	\includegraphics[width=\breite\textwidth]{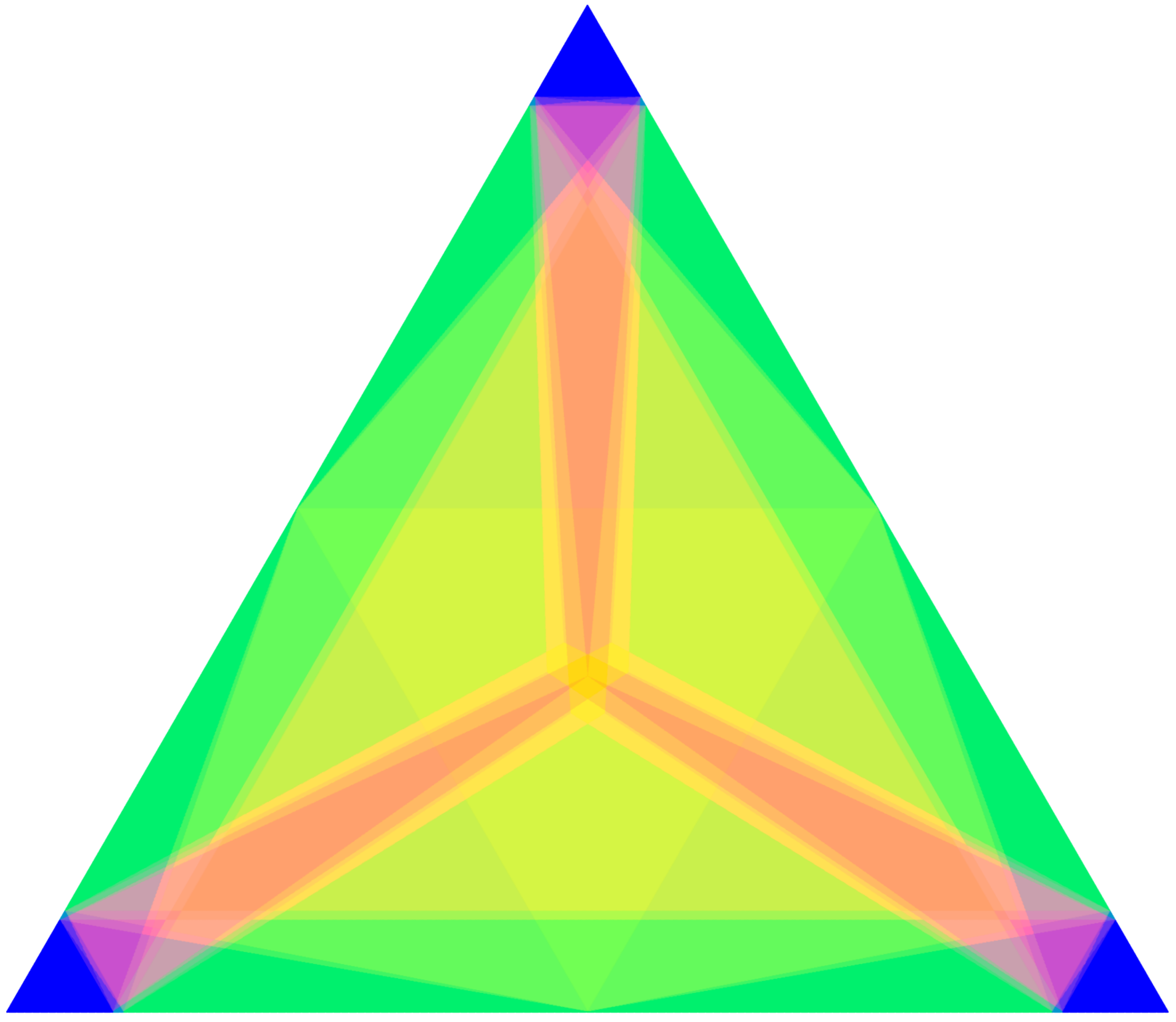}\enspace
	\includegraphics[width=\breite\textwidth]{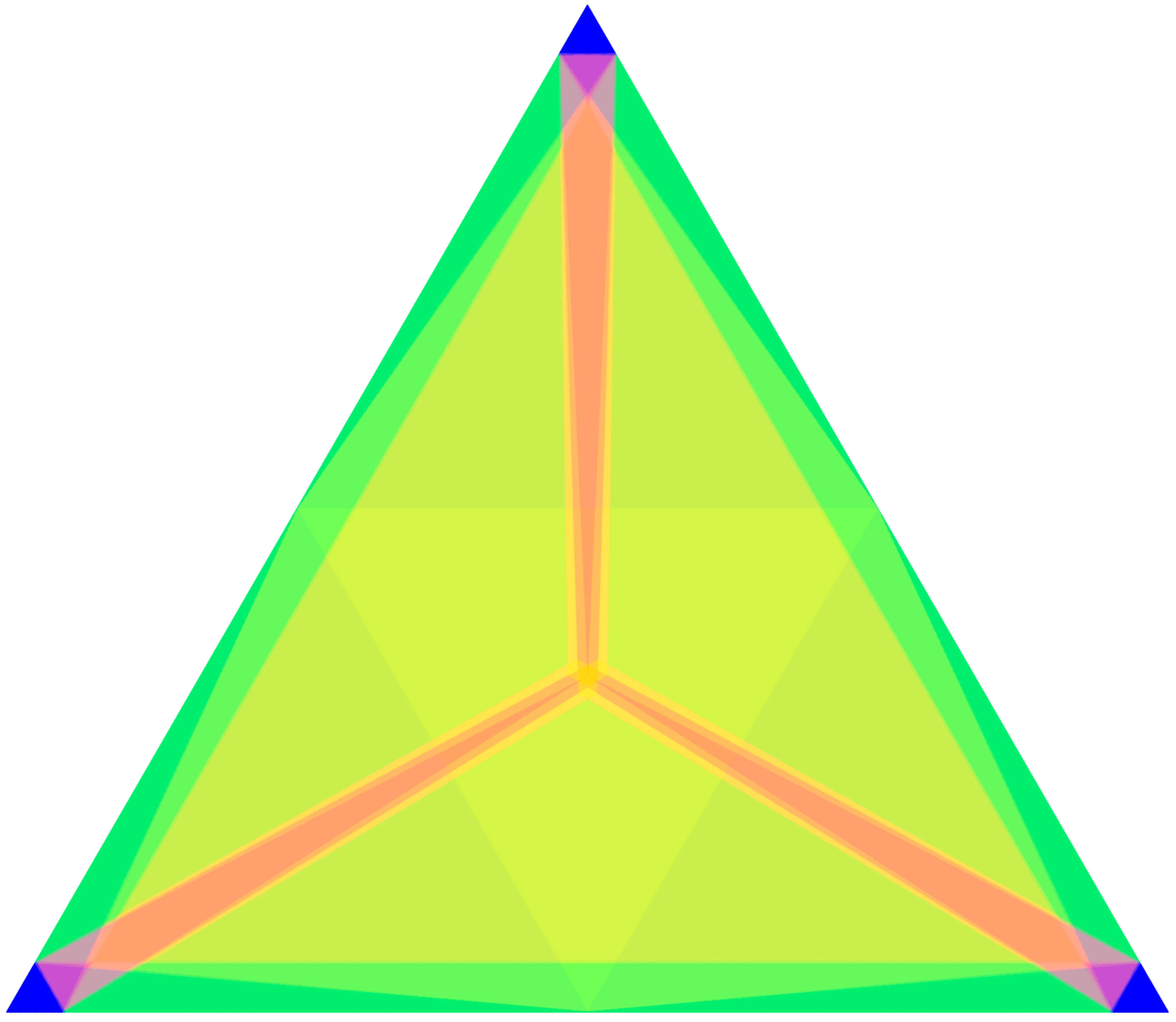}
	\vspace{\titelabstand\textwidth}
	\
	
	{\small
		\makebox[\breite\textwidth][l]{$s=0.90$:}\enspace
		\makebox[\breite\textwidth][l]{$s=0.95$:}}
	
	\vspace{0.44\textwidth}
	
	\includegraphics[width=\breite\textwidth]{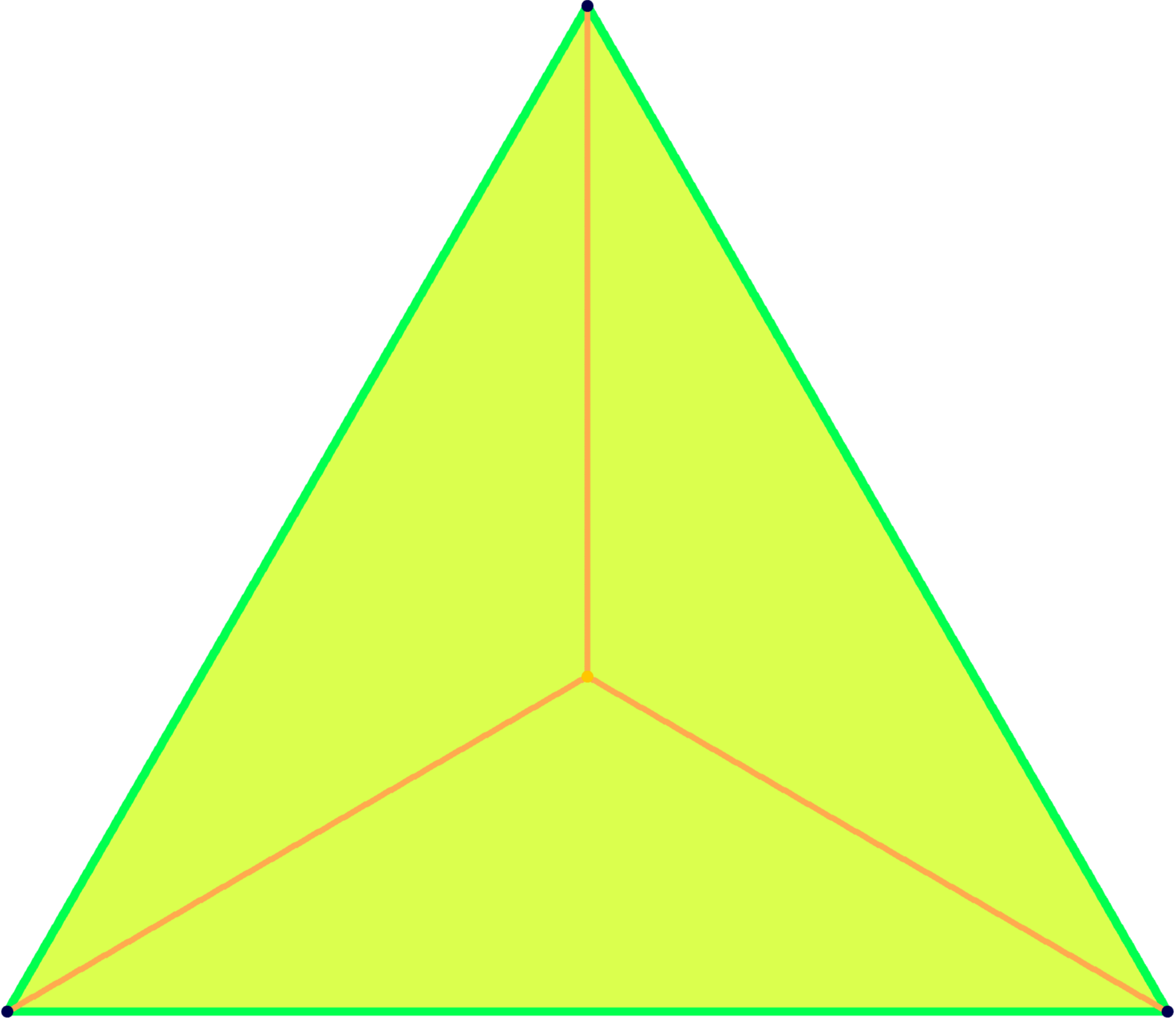}\enspace
	\vspace{-0.445\textwidth}
	\
	
	{\small
		\makebox[\breite\textwidth][l]{$s=1$ (antiplurality):}
	}

	\vspace{\panelabstand\textwidth}
	
	\medskip
	{\captionsize Figure 3 (ctd.): Weighted $s$-scoring committees in Penrose-Banzhaf coloring }
\end{figure}

For instance, the large blue triangles inside the panel for $s=0$, i.e., weighted plurality committees, correspond to $\mathbf{\tilde w}=(1,0,0)$, i.e., dictatorship of one player.
The green midpoints of the simplex's boundary lines represent the equivalence class with $\mathbf{\tilde w}=(1,1,0)$: two players decide symmetrically, the third never makes a difference. 
The simplex's light yellow midpoint reflects $\mathbf{\tilde w}=(1,1,1)$, i.e., three absolutely symmetric players.
The remaining three plurality equivalence classes with reference weights of 
$\mathbf{\tilde w}=(2,1,1)$, $\mathbf{\tilde w}=(2,2,1)$ and $\mathbf{\tilde w}=(3,2,2)$ 
correspond respectively to the purple lines between the boundary midpoints, the dark yellow lines from the simplex's center to the three boundary midpoints, and the residual orange triangles.
Lists of reference weight distributions for other values of $s$ are provided by \citeN{Mayer/Napel:2021:Scoring}.


\section{Concluding remarks}\label{sec:conclusion}

The illustrations in Section~\ref{sec:Bilder} exhibit the hidden beauty of weighted voting in committees.
However, the artistically appealing (at least to us) 
geometry and 
changing colors have substantive implications. They reveal structural properties of collective decision making in politics and economics.

Take, for instance, the large blue triangles in the panel for $s=0$.
As pointed out already, they correspond to dictatorship of one player.
Namely, if some player's voting weight slightly exceeds 50\%~-- because a shareholder has acquired a small majority stake in a corporation, committee seats are awarded in proportion to population shares in an ethnically polarized society with one majority and several large minorities, etc.~-- then all plurality decisions correspond to the top-preference of that player.
This is not the case for many other scoring rules:
consider different levels of $s$ and watch how the blue triangle shrinks from panel to panel.
The shrinkage documents how basing decisions on more than just the top preference over all candidates makes collective choice more `inclusive'.
For instance, adopting Borda's rule instead of taking plurality decisions turns a previous dictator with $\bar w_i$ slightly above 50\% into just a very dominant player.
Player~$i$ can swing the joint decision for many but no longer for all preference configurations under Borda's rule. 
Antiplurality awards dictatorial influence not even to a player who has a perfect monopoly of votes. The player's relative weight of 100\% makes it impossible for the respective worst-ranked candidate to win but with lexicographic tie-breaking relatively few perturbations of the dominant player's preferences alter the winner. 

The changing variety of colors in the panels visualize the findings reported in Figure~\ref{fig:M-shape}: equivalence maps for scoring rules with $0<s<\sfrac{1}{2}$ or $\sfrac{1}{2}<s<1$ involve many more color shades than those for $s=0$, $s=1$, and also $s=\sfrac{1}{2}$.
We can moreover locate the ranges of weights where most of the color changes are concentrated, i.e., where sensitivity to small weight changes is the greatest. 
In these weight regions the incentives to, say, increase one's corporate shareholdings or to try to attract a party switcher are much greater than in monochrome areas.

Multiplicity of colors in a panel also indicates the scope for achieving a particular distribution of influence as an institutional designer who can determine the distribution of weights (the so-called `inverse problem' of voting power; cf.~\citeNP{Kurz:2012}).
Think of a federation of three differently sized states: it may be desirable to make states' voting power a specific function of population sizes~-- e.g., to achieve direct proportionality or proportionality to the square root of population sizes. 
Though perfect symmetry (light yellow) or dictatorship by one state (dark blue) are always feasible, the chances of finding voting weights that achieve the targeted distribution of influence is arguably smaller for, say, $s=1$ with only five equivalence classes than for $s=\sfrac{1}{2}$ with 51. 
There is also a tendency for the distribution of relative voting power to match the underlying distribution of relative voting weights better, the more equivalence classes or colors in our illustrations.
This relates to the so-called `transparency' of a voting rule (cf.~\shortciteNP[Section~7.4]{Kurz/Mayer/Napel:2021:Influence}).

We readily admit that illustrations of voting power in three-player committees that decide between three options have neither the complexity nor the aesthetic qualities of Julia sets or Mandelbrot sets, which have crossed the boundaries between art and science much earlier (see, e.g., \citeNP{Peitgen/Richter:1986}). 
But there is definitely more art and beauty in weighted voting and the resulting voting power than typically meets the eye.


\setlength{\labelsep}{-0.2cm}

%

\end{document}